\documentclass[prd,twocolumn,showpacs]{revtex4}
\usepackage{epsfig}
\usepackage{graphicx}
\usepackage{amsmath}
\usepackage{color}
\usepackage{longtable}

\newcommand{\kB}{k_{\rm B}}
\newcommand{\Tm}{T_{\rm m}}
\newcommand{\Tr}{T_{\rm r}}
\newcommand{\nmax}{n_{\rm max}}

\newcommand{\etaC}{\eta_{\rm C}}

\newcommand{\tauS}{\tau_{\rm S}}
\newcommand{\phaL}{\pha_{\rm L}}

\newcommand{\phaC}{\pha_{\rm C}}

\newcommand{\dphaddeltanu}{\frac{\partial \pha}{\partial (\Delta \nu)}}
\newcommand{\Hethree}{$^3 {\mathrm {He}}$}
\newcommand{\Hefour}{$^4 {\mathrm {He}}$}
\newcommand{\MHz}{{\mathrm {MHz}}}
\newcommand{\GHz}{{\mathrm {GHz}}}
\newcommand{\THz}{{\mathrm {THz}}}

\newcommand{\HI}{H{\sc ~i}}

\newcommand{\HeI}{He{\sc ~i}}
\newcommand{\HeII}{He{\sc ~ii}}
\newcommand{\HeIII}{He{\sc ~iii}}

\newcommand{\LyaHe}{$2^1P^o-1^1S$}
\newcommand{\LyHe}{$n^1P^o-1^1S$}
\newcommand{\InteraHe}{$2^3P^o-1^1S$}
\newcommand{\MquadHe}{$2^3P_2^o-1^1S$}
\newcommand{\InterHe}{$n^3P^o-1^1S$}

\newcommand{\QuadHe}{$n^1D-1^1S$}

\newcommand{\beq}{\begin{equation}}
\newcommand{\eeq}{\end{equation}}
\newcommand{\beqa}{\begin{eqnarray}}
\newcommand{\eeqa}{\end{eqnarray}}

\newcommand{\pha}{\mathcal{N}}
\newcommand{\scl}{\mathcal{L}}
\newcommand{\barr}{\begin{array}}
\newcommand{\earr}{\end{array}}

\begin{document}
\title{Primordial helium recombination III: Thomson scattering, isotope shifts, and cumulative results}
\author{Eric R. Switzer}
\email{switzer@princeton.edu}
\affiliation{Department of Physics, Princeton University, Princeton, New Jersey, 08544, USA}
\author{Christopher M. Hirata}
\email{chirata@sns.ias.edu}
\affiliation{School of Natural Sciences, Institute for Advanced Study, Princeton, New Jersey 08540, USA}
\date{\today}

\begin{abstract}
Upcoming precision measurements of the temperature anisotropy of the cosmic microwave background (CMB) at high multipoles will need to be complemented 
by a more complete understanding of recombination, which determines the damping of anisotropies on these scales.  This is the third in a series of 
papers describing an accurate theory of \HeI\ and \HeII\ recombination.  Here we describe the effect of Thomson scattering, the \Hethree\ isotope 
shift, the contribution of rare decays, collisional processes, and peculiar motion.  These effects are found to be negligible: Thomson and \Hethree\ 
scattering modify the free electron fraction $x_e$ at the level of several$\times 10^{-4}$.  The uncertainty in the \InteraHe\ rate is significant, and for 
conservative estimates gives uncertainties in $x_e$ of order $10^{-3}$.  We describe several convergence tests for the atomic level code and its 
inputs, derive an overall $C_\ell$ error budget, and relate shifts in $x_e(z)$ to the changes in $C_\ell$, which are at the level of $0.5\%$ at $\ell 
=3000$.  Finally, we summarize the main corrections developed thus far.  The remaining uncertainty from known effects is $\sim 0.3\%$ in $x_e$.
\end{abstract}

\maketitle

\section{Introduction}

Cosmological recombination determines the evolution of the free electron fraction as the universe becomes cool enough for bound atoms to form.  
Concurrent with this is a (cosmologically) rapid drop in the Thomson opacity that decouples the motion of the photons from that of the baryons; the 
perturbations in the photons are observable today as the anisotropies of the cosmic microwave background (CMB).  Thus, many properties 
of the measured 
CMB anisotropy are strongly dependent on the free electron history.  The free electron fraction sets the damping scale for temperature 
anisotropies in CMB through Silk damping, and contributes to the acoustic damping scale \cite{1970Ap&SS...7....3S, 1970ApJ...162..815P, 
1987MNRAS.226..655B, 1997Natur.386...37H, 1968ApJ...151..459S, 1997ApJ...479..568H}. These are manifest in the positions of the acoustic peaks in the 
temperature anisotropy power spectrum, and by the suppression of the anisotropy power on small scales.  Because of the large number of modes on the 
sky, the temperature anisotropy power can be measured very accurately, even with small survey coverage.  These modes are the target of a new generation 
of high-precision, small-scale temperature anisotropy experiments \cite{2001PhRvD..63d2001L, 2001PhRvL..86.3475J, 2002MNRAS.334...11A, 
2004AdSpR..34..491T, 2003NewAR..47..939K, 2004SPIE.5498...11R,2004ApJ...600...32K, 2001ApJ...561L...7S, 2002ApJ...571..604N, 2003ApJ...591..556P, 
2002MNRAS.329..890G, 2005MNRAS.363...79G}, and will provide stronger constraints on baryonic and matter fractions, the primordial spectral slope
$n_s$, and its possible scale dependence.

The physics underlying recombination is well-developed and the physical circumstances are simple: the universe is homogeneous, there are no elements 
heavier than Li, and all material is in the gas phase.  The subtleties in the recombination history emerge from both the very high accuracy required, 
and the large number of processes contributing.  Here, rare processes are important because many of the fast allowed processes have reverse processes 
that quickly come to equilibrium, blocking overall progress of the forward direction.  Examples are recombinations directly to the ground state and 
$2^1P^o \rightarrow 1^1S$ decay in \HeI: these are fast but access only certain regions of photon phase space ($h\nu>24.6\,$eV and the 21.2 eV line, 
respectively).  Instead of bringing the ionization state of hydrogen or helium into Saha equilibrium, these processes merely boost the number of photons in 
certain regions of phase space to values far greater than the Planck distribution ${\cal N}=1/(e^{h\nu/\kB T}-1)$ would predict -- until some process 
(such as Sobolev escape in the case of the 21.2 eV line) removes the photons. It is difficult to have a prior notion of what rare processes will 
be significant, and to what extent.  In general, processes that modify the overall recombination history either break the equilibrium in allowed lines 
by removing photons, or provide an entirely independent path to the ground state.  Examples are \HI\ photoionization opacity and two-photon decays, 
respectively.
 
Recently, several corrections to the theory of \HeI\ recombination have been proposed \cite{2006A&A...446...39C, 2005AstL...31..359D, 
2004MNRAS.349..632L, 2006AstL...32..795K}. This is the third of a series of papers describing new effects and remaining uncertainties in 
\HeI\ recombination.  Paper I (Switzer and Hirata astro-ph/0702143) describes the base recombination model, the influence of the feedback of spectral distortions between lines, 
and the effect of continuous opacity from \HI\ photoionization in transport phenomena during \HeI\ recombination.  Paper II (Hirata and Switzer astro-ph/0702144) describes 
absorption of non-thermal radiation in two-photon processes from $n=2$, non-resonant two-photon decay processes from $n>2$, and the effect of finite 
resonance linewidth.  (Henceforth we refer to these as Paper I and Paper II.)  Here, we describe several additional effects that are negligible to 
recombination, summarize the overall magnitude and convergence of effects studied, and give an error budget in free electron fraction and the 
anisotropy power in the CMB.

Because of the large thermal velocities of electrons in the primordial plasma, Thomson scattering results in a very wide redistribution of energies.  A 
typical Thomson scattering will shift the photon's frequency by $\sim 4~\THz$.  For comparison, the Doppler width associated with scattering through atomic 
transitions in \HeI\ is only $\sim 50~\GHz$.  The redistribution width has important consequences for radiative transport within the line subject to Thomson scattering.  
We will see that in the case of the intercombination lines (\InterHe\ in \HeI), it is unlikely for a photon to scatter with an electron as it traverses the width of the 
line that is optically thick to incoherent processes because of the low differential optical depth ($d \tau_e / d \nu$) to Thomson scattering.  
However, a photon can be scattered by an electron as it redshifts far from line center -- as it redshifts, the trajectory integrates sufficient 
differential depth for Thomson scattering to be likely.  Because of the large typical frequency shift from electron scattering, this exchange can 
re-inject the photon onto the blue side of the line, where it has a high probability of redshifting back into the line and being absorbed.  This will 
further excite \HeI\ atoms and inhibit recombination overall.  The allowed lines have sufficient natural width for Thomson scattering to pull photons 
from the region of the line that is optically thick to incoherent processes, which will tend to relax it and acceleration \HeI\ recombination.  Thus in 
\LyaHe, which is most significant for \HeI\ recombination, the effect of Thomson scattering in the far red wings opposes the effect of scattering closer to line center.  

Thomson scattering is a subtle effect because it is tied to the radiation profile on the red side of the line, which is determined by scattering in the wings and \HI\ 
continuous opacity.  It is addressed here by including it in a photon Monte Carlo from Paper I.  We will see, further, that the effect of Thomson scattering is highly suppressed when 
we include feedback between lines.  In the final analysis, including it does not significantly alter the recombination history.

The second main effect examined here is the isotope shift between \Hethree\ and \Hefour, dominated by the nuclear mass difference.  The isotope shift 
between \Hethree\ and \Hefour\ \LyaHe\ is $\nu$(\Hefour)$-\nu$(\Hethree)=$263~\GHz$.  Photons on the red side of $^4$\HeI\ \LyaHe\ can 
scatter off of a \Hethree\ atom and be moved farther into the red damping wing where they are more likely to escape.  However, in the ordinary Sobolev 
theory, this effect is cancelled (on average) by the reverse process where \Hethree\ scatters a photon blueward.  This is because the 
radiation phase space density on the red side of an optically thick line in an expanding background is flat ($\pha$=constant) to a very good 
approximation \cite{1994ApJ...427..603R, 1968ApJ...153....1P, 1957SvA.....1..678S}.  As we saw in Paper I, continuous opacity from \HI\ 
produces a gradient in the radiation phase space density, as progressively more photons that escape the line are absorbed, moving redward.  Thus, once 
the \HI\ population becomes significant at $z<2200$, the radiation phase space density on the red side of the line is no longer flat.  
This means that, on average, more photons can be scattered redward (because the ``pool" is larger) than blueward, and the escape probability increases.  
We find that this causes a small modification to the free electron fraction, $|\Delta x_e| < 2.5 \times 10^{-4}$, peaking at $z\sim 1900$.

In Paper I we calculate the modification of rates in the \InterHe\ and \QuadHe\ series by continuous opacity.  Here we describe additional sources of 
error in the recombination calculation related to rare processes and a breakdown of the relative contributions of the rare processes.  There is 
considerable disagreement on the spontaneous \InteraHe\ rate in the literature \cite{1977PhRvA..15..154L, 1969ApJ...157..459D, 1978JPhB...11L.391L}.  
This is a significant systematic overall in \HeI\ recombination, resulting in a maximum uncertainty at $z\sim 1900$, of roughly $|\Delta x_e| < 1 
\times 10^{-3}$.  The \QuadHe\ series and \InterHe\ for $n \ge 3$ are found to give negligible modifications to the \HeI\ recombination history.

In contrast to the situation in the interstellar medium, collisional processes during cosmological recombination are subdominant
because of 
the low baryon/photon ratio.  During \HeI\ recombination, the highly excited states are also so close to equilibrium that electron collision-induced 
transitions between bound and free states cannot do anything (Sec.~\ref{ss:collisionsmain}).  Two remaining possibilities for collisional processes 
that modify the \HeI\ recombination history are charge transfer reactions with \HI\ (through a collision, a neutral hydrogen atom transfers its 
electron to singly-ionized helium atom) and collisional de-excitations to the ground state (where collisions de-excite and atom to its ground state
without the emission of photon that will further excite that transition) \cite{1998ApJ...509....1S, 1990JPhB...23.4321S}.  
In general, charge transfer requires an \HI\ population.  Continuous opacity 
(described in Paper I) causes \HeI\ recombination to finish as soon as even a small amount of \HI\ is present, and we will see that collisional charge transfer 
plays a subdominant role.  In Sec.~\ref{ss:collisionsmain}, we also show that the collisional de-excitations to the ground state are negligible from 
the combination of temperature, free electron fraction and collision strengths.

Throughout the series of papers we consider transport processes during \HeI\ recombination that depend explicitly on the radiation profile around 
the line and the line profile.  This is in contrast to Sobolev methods for the escape probability, which do not depend on the line profile.  A significant concern
is that we have ignored the correlated, peculiar motions in the gas.  In astrophysical systems, microturbulance leads to broadening of the line profile in
addition to the broadening from thermal motion.  Indeed, correlated motions in the recombination plasma are of the same order as the thermal velocities.
In Sec.~\ref{ss:peculiar}, we argue that regions of correlated motion are much larger than the distances a photon traverses as it scatters through the line.  Thus,
to a very good approximation, the recombination picture is unchanged.

We propose three main corrections to the standard \HeI\ recombination model based on the studies in this series of papers: (1) inclusion of feedback 
between allowed lines from redshifting of the spectral distortion from $n+1$ to $n$, (2) inclusion of rates from intercombination processes 
\cite{2005AstL...31..359D}, and (3) treatment of \HI\ continuous opacity in transport within \LyaHe\ and \InteraHe.  These corrections were addressed 
in Paper I, and the results from Paper II and this paper have little effect.  (Note that, throughout, we have performed very detailed analyses of the effects that we find to be 
important, or in cases where only a detailed analysis can determine whether the effect is important.  Conversely, we have not investigated 
processes in detail if a simple order-of-magnitude argument shows that they are unimportant.  The approximate treatments are indicated with a 
$\sim$ sign in front of the magnitude of the effect in the summary in Table~\ref{tab:table1}).

In this paper we break the discussion of effects into several sections.  In Sec.~\ref{sec:thom}, we describe the effect of Thomson scattering in the 
$n=2$ allowed and intercombination lines.  In Sec.~\ref{sec:heliumthree}, we treat the isotope shift of \Hethree\ relative to \Hefour\ in the 
transition \LyaHe.  Next, in Sec.~\ref{sec:additionalsys}, we study several additional uncertainties: treatment of intercombination 
and quadrupole lines (Sec.~\ref{ss:rare}); collisional processes (Sec.~\ref{ss:collisionsmain}); peculiar velocities (Sec.~\ref{ss:peculiar}); and 
convergence of the numerical code (Sec.~\ref{ss:convergence}).  In Sec.~\ref{sec:CMBaniso}, we relate modifications in the \HeI\ recombination history 
to modifications in the 
temperature-temperature (TT), polarization-temperature (TE), and polarization-polarization (EE) anisotropy power as a function of multipole.  In 
Sec.~\ref{sec:summary}, we summarize the contributions to $x_e(z)$ and anisotropy error budget associated with effects investigated in this series
and describe the prospects for including these effects in a fast multilevel recombination code for CMB codes.  
Appendix~\ref{sec:appcharele} describes an explicit form for the electron scattering kernel, and Appendix~\ref{app:cpt} describes cosmological perturbation
theory relevant for the helium peculiar velocity.

\section{Thomson Scattering}
\label{sec:thom}

Consideration of the effect of Thomson scattering in line transport during \HeI\ recombination is well-motivated because the bulk of electrons have not 
recombined, and their large thermal velocity dispersion relative to \Hefour\ atoms yields new transport behavior.  In this section we analyze this 
behavior and relevance to the \HeI\ recombination history.  Electron scattering should also be folded into radiative transport in \HI\ recombination;
this is beyond the scope of discussion here and is deferred to later work.  This is a harder problem because of the high optical depth, broad linewidth and partial 
redistribution in the \HI\ $2p \leftrightarrow 1s$ system, as well as the more stringent accuracy requirements.  In particular the very high optical 
depth (of order $10^9$) may require a Fokker-Planck or hybrid approach to the problem instead of the Monte Carlo methods applied here.

As in Paper I, a photon Monte Carlo will be the workhorse for calculating the modified escape probability in the \HeI\ \InterHe, \QuadHe, and \LyaHe\ 
lines.  This is described in Sec.~\ref{ss:eMC}.  The physics of the scattering kernel is described in Sec.~\ref{ss:ekernels} and 
Appendix~\ref{sec:appcharele}.  We also develop analytic methods (Sec.~\ref{ss:qintegral}) based on this kernel to check the Monte Carlo with forbidden 
lines in restricted cases.  In forbidden lines, the situation is simplified relative to \LyaHe\ because the photons are completely distributed across a 
line which is narrow compared to the thermal width of electrons, so that accurate analytic transport solutions are feasible.  We find that Thomson 
scattering is negligible overall due to cancellation of several competing effects and suppression from feedback of spectral distortion between lines.

\subsection{Physical setting}
\label{ss:physthom}

\begin{figure}
\includegraphics[width=3.2in]{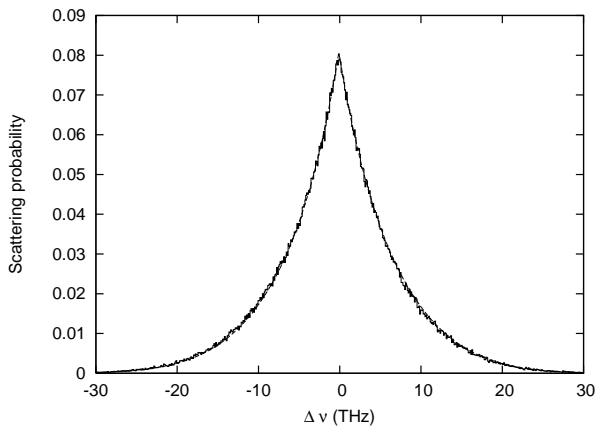}
\caption{\label{figs:escat} The electron scattering kernel at $z=2500$ spans several THz, allowing photons in the far red side of the line to be scattered 
to frequencies significantly above the line center frequency.  Rather than escaping on the red side of the line, as in ordinary Sobolev escape, the photon 
can then be absorbed (and likely be re-emitted) by incoherent processes in the line before it can escape, thus reducing its escape probability.  Because 
the radiation phase space density is higher on the red side of the line, a photon is more likely to scatter from below to above the line frequency, thus 
decreasing the overall escape probability.  Once continuum opacity becomes significant in the allowed lines, this can also remove trapped photons.}
\end{figure}

In Paper I, we considered line radiative transport in a homogeneous gas on an expanding background subject to coherent scattering through the line, 
incoherent processes in the line, and \HI\ photoionization.  Here, we extend this picture to include Thomson scattering from a thermal electron 
distribution with temperature $\Tm$ (though this is nearly identical to $\Tr$ during \HeI\ recombination).  Depending on the electron's velocity, a 
photon will be widely redistributed because the characteristic scattering width for a photon of frequency $\nu_0$ is $\sim \nu_0 \sqrt{\kB\Tm / (m_e 
c^2)}$.  This redistribution is shown at $z=2500$ in Fig.~\ref{figs:escat}, and is much larger than the Doppler width of the \Hefour\ line, $\sim 
50$~GHz.  Fundamentally, the only reason for solving the transport problem is to find the probability $P_{\rm esc}$ that a photon emitted by an atom 
will be absorbed through incoherent processes in another atom before it can escape the line.  That is, the full matrix of probabilities (that a photon 
which has just been emitted or scattered will be emitted or scattered by given process before it escapes) is immaterial, aside from its contribution to 
$P_{\rm esc}$, as in Paper I.  We will consider the modification to escape probabilities from Thomson scattering in \LyHe\ ($n<6$), \InterHe\ ($n<5$), 
and \QuadHe\ ($n<6$).

\begin{figure}
\includegraphics[width=3.2in]{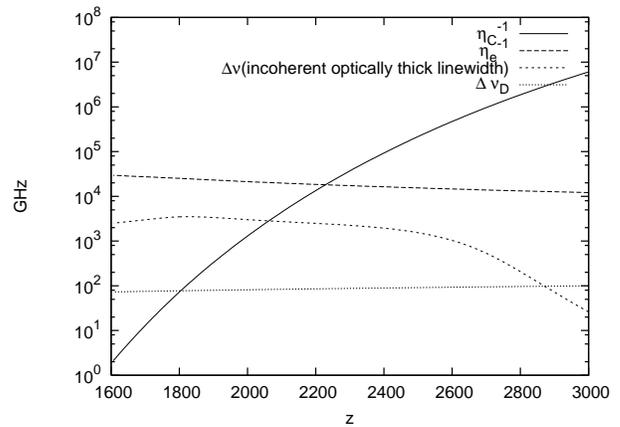}
\caption{\label{figs:electronscattering} The inverse differential optical depth due to Thomson scattering of free electrons ($\eta_e^{-1}$), compared to the inverse 
differential depth to \HI\ photoionization ($\etaC^{-1}$), the line Doppler width ($\Delta \nu_{\rm D}$), and the optically thick linewidth to incoherent processes 
through \LyaHe.  Here we can see that a photon trapped in a line is relatively unlikely to interact with an electron before it escapes. For $z<2500$, 
electron scattering acts on scales only roughly an order of magnitude larger than the optically thick linewidth to incoherent processes.  The behavior 
has three regimes for \LyaHe: (1) for $z> 2200$ electron scattering influences transport by acting over large scales, retarding recombination by 
injecting photons to optically thick regions on average; (2) \HI\ continuous opacity suppresses the phase space density on large scales: when electron 
scattering occurs, it is much closer to the line's center and removes trapped photons, accelerating \HeI\ recombination; and (3) $z < 1900$ where \HI\ 
continuous opacity acts within the line and dominates most of the behavior.}
\end{figure}

Electron scattering shares some, but not all of the features of continuous opacity from neutral hydrogen, developed in 
Paper I.  Like continuous opacity in \HI\, electron scattering presents a differential optical depth approximately flat in frequency,
\beq \eta_e = \frac{n_{\rm H} x_e \sigma_{\rm T} c }{H(z) \nu_{\rm line}}. \label{eqn:etaedef} \eeq
Electron scattering would naively be expected to become important if $\eta_e^{-1}$ starts to fall within frequency width of the line that is optically 
thick to incoherent processes, $\Delta\nu_{\rm line}$.  In Paper I, this allowed a small \HI\ population to remove photons from the line core, so that
they can no longer further excite the line; thus increasing the escape probability.  The relevant comparison for electron scattering is plotted in 
Fig.~\ref{figs:electronscattering}, which makes it clear that electron scattering acts only on frequency scales an order of magnitude larger than the 
incoherent optically thick width of the line.  Therefore one might expect an accelerating effect of order $\sim 10$\% due to electron scattering.  
In reality this number will be smaller than $\sim 10$\% because the Thomson opacity -- unlike \HI\ photoionization -- is from scattering, not absorption.  
Some of the scattered photons will re-enter the line, and some photons that are on the red side of the line and would otherwise have escaped will be 
scattered into the line, or to its blue side (where they redshift back into the line).  If there are more photons on the red side of the line than the blue, more 
photons are scattered bluewards than redwards on average, and the escape probability can be decreased.  In particular, this means that it is also possible 
for Thomson scattering to {\em delay} \HeI\ recombination.

\subsection{Monte Carlo simulation of line transport with Thomson scattering}
\label{ss:eMC}

In this section, we use a Monte Carlo method to estimate the escape probability for a photon trapped in \HeI\ lines.  For reasons described in Paper I, 
a Monte Carlo is a useful procedure to account for the numerous effects active in the radiative transport within lines, despite the computational cost.  
In particular, the large width of the electron scattering kernel compared to the width of the \HeI\ lines suggests that a Fokker-Planck approach is not 
well suited to evaluating electron scattering.  (The Fokker-Planck operator treats transport of numerous scattering events as a continuous process, and is
accurate if the line profile varies slowly over this range.  In electron scattering, a photon in the far red can be scattered to blue, and this assumption is 
invalid.   We will consider an alternative analytical approach in Sec.~\ref{ss:qintegral} that is useful for testing the Monte Carlo.)

Kinematically, the frequency shift in a photon scattering from an electron in the non-relativistic limit is
\beq \Delta \nu = (f_\parallel + \alpha)(1- \cos\chi) - f_\perp \sin\chi, \label{eqn:escatfunc} \eeq
where ${\mathbf f}= \nu_{0} {\mathbf v}/c$ is thermally distributed ($\parallel$ and $\perp$ denote the components parallel to the direction of 
propagation of the incident photon, and perpendicular to it but in the plane of scattering, respectively), for the photon frequency $\nu_{0}$, $\alpha 
= h \nu_{0}^2/(m_e c^2)$, and $\chi$ is the scattering angle between the direction of propagation of the incoming/outgoing photons.  The recoil term $\alpha$ 
is from differences in the electron kinetic energy and can be dropped (it is of order $200~\GHz$, small compared to a frequency shift of several THz typical in the 
exchange due to the Doppler shift).  This is nearly identical to the case of coherent line scattering, except that the scattering cross section is flat 
in frequency and struck atom's velocity is not conditioned on the incoming photon frequency.  This can be trivially added to the Monte Carlo developed in 
Paper I by adding an additional scattering process with differential depth $\eta_e$, from Eq.~(\ref{eqn:etaedef}).

As in Paper I, we simulate the escape probability over a grid in $\{ x_{\rm HeI}, z \}$ with continuous opacity derived from \HI\ populations in Saha 
equilibrium (accurate until the end of \HeI\ recombination, where it ceases to matter anyway, once \HeI\ has almost fully recombined).  The grid is 
taken over 11 linearly-spaced points in redshift spanning $z=1400$ to $z=3000$, and 21-logarithmically spaced points in $x_{\rm HeI}$ from $2 \times 
10^{-5}$ to $0.08$.  This is log-interpolated along $x_{\rm HeI}$ and $P_{\rm esc}$, and linearly interpolated along $z$ in the level code.

\begin{figure}
\includegraphics[width=3.2in]{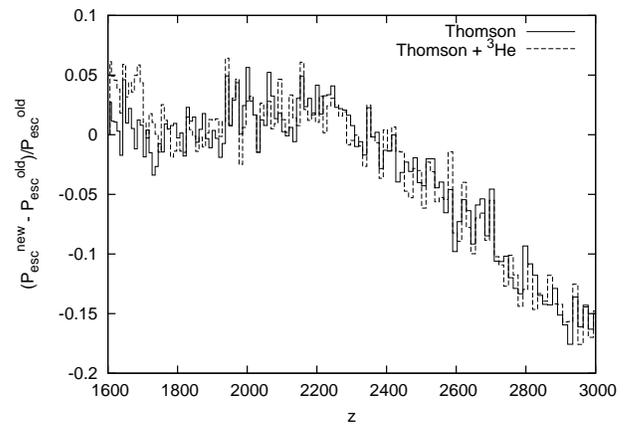}
\caption{\label{figs:escat_prob} The fractional modification to the \LyaHe\ escape probability due to electron scattering and \Hethree\ scattering.  The effect of electron scattering can be 
split into two regimes: 1) for $z>2300$ where most of the effect from electron scattering is due to modification to transport far in the wings and the escape 
probability is decreased, 2) for $z<2300$, where much of radiation in the red wing is absorbed through \HI\ photoionization, and electron scattering begins to
eject more photons from within the line, on average.  Thus, for $1600 < z < 2300$ electron scattering slightly increases the escape probability.  \Hethree\ can
be seen to be a small effect--also note that both effects are at the level of the noise in the MC measurement.  These required 3 days across $50\times3$~GHz
nodes, so improved statistics would require significant computing time.} 
\end{figure}

The results of the Monte Carlo calculation of the escape probability for a typical recombination history are shown in Fig.~\ref{figs:escat_history} for 
\InteraHe\ and in Fig.~\ref{figs:escat_prob} for \LyaHe.  (Fig.~\ref{figs:escat_history} also has results from an analytic method developed in Sec.~\ref{ss:qintegral} 
for the escape probabilities in \InteraHe\ for $z>2000$.  Fig.~\ref{figs:escat_prob} includes Monte Carlo results for \Hethree\ scattering, developed in 
Sec.~\ref{sec:heliumthree}.)  Fig.~\ref{figs:escat_history} indicates that Thomson scattering leads to a reduced escape probability for $z>2100$ in \InteraHe\ 
and \LyaHe, as photons that redshift out of the line are Thomson scatttered back on onto it, where they are less likely to escape.  Once \HI\ opacity becomes 
important, fewer photons are able to travel far enough to be Thomson scattered, and are, instead, more likely to photoionize an \HI\ atom.

\subsection{The electron scattering kernel}
\label{ss:ekernels}
 
The electron scattering redistribution kernel has accurate approximate expressions over a wide range of physical scales \cite{2000ApJ...543...28S}. In 
the non-relativistic limit considered here, the Thomson scattering kernel with dipolar angular distribution \cite{1967ApJ...150L..57H} is adequate.  
We will actually use both the kernel and its characteristic function, which can be obtained in closed-form from the angular dependence 
(Eq.~\ref{eqn:escatfunc}) as follows.  The electron velocity components $f_\parallel$ and $f_\perp$ are normally 
distributed with variance
$\sigma_D^2=\nu_0^2 \kB\Tm/ (m_e c^2)$.  Thus, for a fixed scattering angle $\chi$, the change in frequency is 
equal to the sum of two normal distributions centered around zero, and we write the variance of the photon frequency for a given $\chi$,
\beq \sigma_\nu^2(\chi) = 2 \frac{k_B T_e}{m_e c^2} \nu_0^2 (1-\cos\chi). \eeq
Integration against the angular redistribution probability gives the redistribution kernel
\beq P(\Delta \nu) = \left \langle \frac{1}{\sqrt{2 \pi} \sigma_\nu(\chi)} \exp \left \{ - \frac{\Delta \nu^2}{2 \sigma_\nu(\chi)^2} \right \} \right \rangle_{\chi}. \eeq
The dipole angular redistribution for electron scattering is
\beq P_{\rm dipole}(\chi) \,d \chi = \frac{3}{8} (1 + \cos^2\chi) \sin\chi\, d \chi. \eeq
The kernel exists in the literature \cite{1967ApJ...150L..57H} in terms of Gaussian and error functions.
In Fig.~\ref{figs:escat} we compare the kernel (in physical frequencies $\Delta \nu$) for dipole scattering with a with a Monte Carlo 
calculation.  For the analytic work of Sec.~\ref{ss:qintegral} (needed to test the Monte Carlo) it is more 
convenient to work in the Fourier domain, so instead of 
$P(\Delta \nu)$ \cite{1967ApJ...150L..57H} we will concentrate on the characteristic function of the photon $\Delta \nu$ distribution,
\beq \varpi(k) = \langle e^{ik\Delta \nu} \rangle_{\Delta \nu}
\equiv \int_{-\infty}^\infty e^{ik\Delta \nu} P(\Delta\nu)\,d\Delta\nu.
\eeq
This has a closed form solution for the dipole angular redistribution function that is given Appendix~\ref{sec:appcharele}.

\subsection{An approximation to transport in Doppler-width dominated lines with electron scattering}
\label{ss:qintegral}

In the limit that the linewidth is small compared to the characteristic redistribution width for electron scattering and the \HI\ photoionization 
opacity within the line ($\eta_{\rm C}\Delta\nu_{\rm line}$) is small, we can derive an approximate solution for the modification to the escape 
probability with complete redistribution.  This is a reasonable approximation in the case of the intercombination lines in \HeI\ recombination, where 
the characteristic width is $\approx 10^2\,$GHz and the electron scattering redistribution width is several THz.

To find the escape probability using an analytic method, one generally solves for the difference between the phase space density of radiation in 
equilibrium with the line and the actual radiation phase space density integrated across the line profile, $\phaL - \bar \pha$ (see Paper I).  Taking 
$\Delta \nu$ relative to the line center ($\nu = \Delta \nu+ \nu_{\rm line}$), the transport equation with electron scattering with complete 
redistribution, continuous opacity and electron scattering is
\beqa \dphaddeltanu &=& \eta_{\rm C} [\pha(\Delta \nu) - \phaC] + \tauS \phi(\nu) [\pha(\Delta \nu) - \phaL] \nonumber \\ &&+ \eta_e \biggl [ 
\pha(\Delta \nu) \nonumber \\ &&- \int p(\Delta \nu - \Delta \nu') \pha(\Delta \nu') d \Delta \nu' \biggr ].
\label{eq:eseq1}
\eeqa
The approximation we take here is that $\phi(\Delta \nu) \rightarrow \delta(\Delta \nu)$, which reduces this equation to
\beqa \dphaddeltanu &=& \eta_{\rm C} [\pha(\Delta \nu) - \phaC] -I\delta(\Delta\nu) \nonumber \\ &+& \eta_e \biggl [ 
\pha(\Delta \nu) \nonumber \\ &-& \int p(\Delta \nu - \Delta \nu') \pha(\Delta \nu') d \Delta \nu' \biggr ],
\label{eq:eseq}
\eeqa
where
\beq
I = \pha(-\epsilon)-\pha(+\epsilon)
\label{eq:idef}
\eeq
is the jump across the line.  If we consider Eq.~(\ref{eq:eseq1}) in the immediate vicinity of the line, we can see that the right-hand side contains a 
delta function at $\Delta\nu=0$, and hence there is a jump in $\pha$ at this value.  The jump condition is obtained by considering the integral 
$\Phi(\Delta\nu)=\int_{-\epsilon}^{\Delta\nu}\phi(\Delta\nu')\,d\Delta\nu'$, which varies from $0$ at $\Delta\nu=-\epsilon$ to $1$ at 
$\Delta\nu=+\epsilon$.  The relevant terms in Eq.~(\ref{eq:eseq}) are
\beq
\frac{\partial\pha}{\partial\Phi} = \tauS (\pha-\phaL),
\eeq
so that $\pha-\phaL\propto e^{\tauS\Phi}$.  This has the solution
\beq
\phaL - \pha(-\epsilon) = [ \phaL - \pha(+\epsilon)] e^{-\tauS},
\label{eq:jump}
\eeq
and the phase-space density averaged across the line is
\beq
\bar\pha =\int_0^1\left\{ \phaL - [\phaL - \pha(+\epsilon)]e^{-\tauS(1-\Phi)} \right\} d\Phi = \phaL - \frac I{\tau_{\rm S}}.
\label{eq:pha-I}
\eeq

\begin{figure}
\includegraphics[width=3.2in]{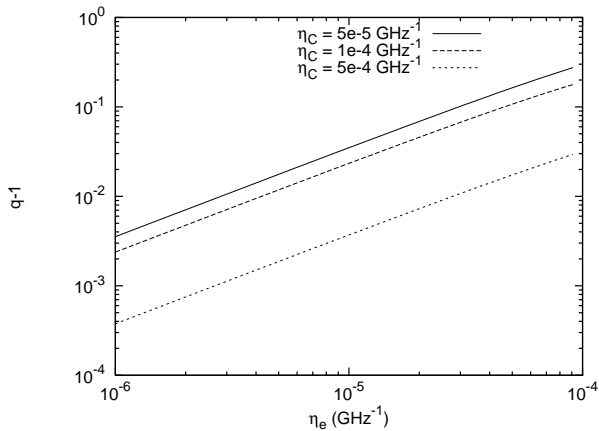}
\caption{\label{figs:qintegral}  The quantity $q-1$ (defined in Eq.~(\ref{eqn:qintegral})), which quantifies the departure from the Sobolev theory due to electron 
scattering as a function of the electron scattering differential opacity, for several values of the continuum depth for \InteraHe\ at $z=2500$, which sets the electron 
temperature.}
\end{figure}

\begin{figure}
\includegraphics[width=3.2in]{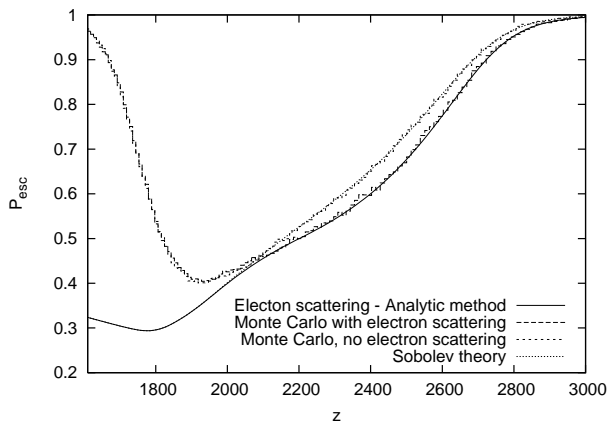}
\caption{\label{figs:escat_history}  A comparison of analytic and Monte Carlo treatments of transport in \InteraHe\ which give a modified escape probability 
due to electron scattering.  This particular trajectory of $x_{\rm HeI}$ is for the recombination history developed in Paper I.  The analytic description of 
electron scattering in narrow lines with complete redistribution of Eq.~(\ref{eqn:pescwithe}) agrees well with Monte Carlo results at early times.  Once 
continuous opacity becomes important within the line, the Monte Carlo gives a dramatic increase in the escape probability.  Built in to the analytic method, 
however is the assumption that the line is indefinitely narrow.  Thus, the analytic method breaks down $z< 2100$ and approaches the Sobolev theory. 
(The history without electron scattering is also consistent with the analytic method for complete redistribution and continuous opacity developed in Paper I.)}
\end{figure}

In order to proceed further, in particular to evaluate $I$, we need one more relation between $\pha(-\epsilon)$ and $\pha(+\epsilon)$.  This can be 
obtained by transforming Eq.~(\ref{eq:eseq}) into the Fourier domain:
\beq i k \scl(k)  = \eta_{\rm C} \scl(k) + \eta_e[1 - \varpi(k)] \scl(k) - I, \eeq
where
\beq \scl(k) = \int \left [ \pha (\Delta \nu) -\phaC \right ] e^{ik(\Delta \nu)} d(\Delta \nu). \eeq
This has the simple solution
\beq
\scl(k) = \frac I {\eta_{\rm C}+\eta_e[1 - \varpi(k)]-ik}.
\label{eq:scl}
\eeq
The phase space density averaged over the jump is then the inverse Fourier transform,
\beq
\frac12[\pha(+\epsilon) + \pha(-\epsilon)] = \phaC + \frac{1}{2\pi}PP\int_{-\infty}^{\infty} \scl(k)\,dk,
\eeq
where $PP$ denotes the principal part.  Using Eq.~(\ref{eq:scl}), this can be re-written as
\beq \pha(+\epsilon) + \pha(-\epsilon) = 2 \phaC + I q,
\label{eq:q-use}
\eeq
where
\beqa
q &=& \frac 1{\pi}PP\int_{-\infty}^{\infty} \frac I {\eta_{\rm C}+\eta_e[1 - \varpi(k)]-ik} dk
\nonumber \\
&=& \frac{2}{\pi} \int_0^\infty \frac{\etaC + \eta_e [1-\varpi(k)]}{k^2 + \{ \etaC + \eta_e [1-\varpi(k)] \}^2} dk. \label{eqn:qintegral}
\eeqa
We can solve for $I$ by algebraically combining Eq.~(\ref{eq:q-use}) with the definition of $I$ (Eq.~\ref{eq:idef}) and with Eq.~(\ref{eq:jump}).  This gives
\beqa
I &=& \frac{2(\phaL-\phaC)(1-e^{-\tau_S})}{1+q+(1-q)e^{-\tau_S}}
\nonumber \\
&=& (\phaL-\phaC) \frac{2}{q+\coth(\tau_S/2)}.
\eeqa
From Eq.~(\ref{eq:pha-I}) we find
\beq \phaL - \bar \pha \approx \frac{1}{\tau_S} (\phaL - \phaC) \frac{2}{q+\coth(\tau_S/2)}, \eeq
giving the modified escape probability
\beq P_{\rm esc} = \frac{2}{\tauS[q+\coth(\tau_S/2)]}. \label{eqn:pescwithe} \eeq
Taking the limit that $\eta_e \rightarrow 0$, one can see that $q\rightarrow 1$, regardless of the continuum opacity, $\etaC$.  Through the use of 
hyperbolic function identities the $q=1$ case can be shown to give the Sobolev result, $P_{\rm esc}=(1-e^{-\tauS})/\tauS$.
This is by construction of the approximation -- for an indefinitely thin line, finite continuous opacity can not affect transport ``within" the line.  
We evaluate Eq.~(\ref{eqn:qintegral}) for $q$ using a 15-point Gauss-Kronrod rule, using the closed form 
solution for $\varpi(k)$ from Appendix~\ref{sec:appcharele}.  The value $q$ is shown as a function of $\eta_e$ for several choices of the 
continuum differential optical depth in Fig.~\ref{figs:qintegral}.  We can also use Eq.~(\ref{eqn:pescwithe}) to find the modified escape probability 
for a typical recombination history and compare with Monte Carlo results.  This is shown in Fig.~\ref{figs:escat_history}.

Note that the analytic method derived here is only applicable when continuous opacity does not act within the optically thick width of the line.  In 
principle, if the intercombination lines remain narrow compared to $\eta_{\rm C}^{-1}$ until $z < 1800$, then \HeI\ will have nearly finished 
recombination and the analytic method of Eq.~(\ref{eqn:pescwithe}) would be appropriate (even though it is technically incorrect after that point, it 
ceases to matter).
In a comparison to Monte Carlo results in Fig.~\ref{figs:escat_history}, we see that there is significant departure from the assumptions built into 
Eq.~(\ref{eqn:pescwithe}) starting at $z \sim 2000$.  Thus in the level code we use an interpolated grid of probabilities from the Monte Carlo, and the 
method here should only lend confidence to the Monte Carlo result for $z>2000$, where electron scattering matters most.

\section{\Hethree\ and the isotope shift}
\label{sec:heliumthree}

Up to now, we have considered the \HeI\ lines to have Voigt profiles.  This is not quite correct, because the transition frequencies in \Hethree\ are 
shifted slightly to the red of than analogous transition in \Hefour.  In the usual Sobolev approximation this does not have any effect because the line profile is irrelevant.  
However, when \HI\ opacity is included, it is possible that splitting the line into two pieces (one for \Hethree\ and one for \Hefour) would speed up 
recombination.  This could occur either if photons redshifting out of the \Hefour\ line are absorbed by \HI\ before reaching the \Hethree\ line, or if 
scattering in the \Hethree\ line can transport photons farther into the red wing of the \Hefour\ line where they are more likely to escape.  In 
practice the effect of \Hethree\ in radiative transport is more subtle than these simple arguments because the \Hethree\ line overlaps both the red 
damping tail and the Doppler core of \Hefour.  This section describes how \Hethree\ scattering is included in the photon Monte Carlo simulation.  We 
argue that the overall contribution to the escape probability is negligible.

Among the \LyHe\ series, the \LyaHe\ line is the primary contribution to the \HeI\ recombination rate. We will consider the modification of radiative 
transport in \LyaHe\ caused by \Hethree\ scattering photons from the main \Hefour\ line.  If the modification to \LyaHe\ is negligible, then it may be 
assumed that the higher order \LyHe\ series lines have lesser contribution to \HeI\ recombination.  (Even if \Hethree\ led to a larger modification in 
the escape probability for $n>2$, the levels are so sparsely populated that their overall contribution is subdominant.)  The isotope shift is 
inconsequential to transport in intercombination and quadrupole lines because the analogous \Hethree\ feature is very optically thin.

The \LyaHe\ line has an isotope shift \cite{2006CaJPh..84...83M} of
\beq \nu(^3{\rm He})-\nu(^4{\rm He}) = -263~\GHz. \eeq
This difference is approximately four thermal widths of the \Hefour\ line during most of \HeI\ recombination, so both the separation of the lines and the 
larger Doppler width associated with \Hethree\ must be considered.  Technically the $2^1P^o$ level in \Hethree\ also has hyperfine structure, with 
a splitting of $20.8\;\MHz$ \cite{2006CaJPh..84...83M} between $F=1/2$ and $F=3/2$ levels, but this is well inside the Doppler or natural width of the \Hethree\ line and so can 
be neglected.  Throughout, we use an abundance ratio by number of \Hethree:\Hefour\ of $f_4(^3{\rm He})=1.1\times 10^{-5}/0.079=1.4\times 10^{-4}$
\cite{2007ApJS..170..377S}.

Reciprocity holds to a very good approximation in \Hethree\ scattering, i.e. the rate for scattering a photon from frequency $\nu_1$ to $\nu_2$ is 
equal to the reverse rate from $\nu_2$ to $\nu_1$.  (As discussed in Paper II, this is violated if $h\nu_1-h\nu_2$ is of order $\kB\Tm$, but for the 
linewidths relevant in helium recombination this does not happen.) For a radiation field whose phase space density is constant across the width of the 
\Hethree\ line, scattering by \Hethree\ cannot have any effect.  On average, for each photon scattered to the red side of the \Hethree\ line, there is one 
scattered to the blue side of the \Hethree\ line.  In \HeI\ recombination without continuous opacity from \HI, the radiation on the red side of the \Hefour\ 
\LyaHe\ line is very flat over the width of the \Hethree\ line \cite{1994ApJ...427..603R, 1968ApJ...153....1P, 1957SvA.....1..678S}.  Similarly if the 
\Hethree\ and \Hefour\ level occupation probabilities are equal and there is no \HI\ opacity, then the radiation phase space density produced in the 
optically thick \Hefour\ line is already in equilibrium with the excitation temperature of the \Hethree.  Therefore without \HI\ continuum opacity the 
\Hethree\ can have no effect.  However, continuum opacity from \HI\ becomes significant for $z<2200$, and produces a gradient in the radiation phase 
space density on the red side of \Hefour\ \LyaHe.  The interaction of partial redistribution through \Hefour\ and \Hethree\ (though only coherent 
scattering in \Hethree\ is significant) and \HI\ continuous opacity is best dealt with numerically.

The Monte Carlo provides a convenient way to include scattering through \Hethree\ in the radiative transport in the \Hefour\ \LyaHe\ line.  Because we 
take an identical occupation history for \Hethree\ and \Hefour, and the isotope shift is small compared to line-center frequency, the optical depth is 
well-approximated as $\tau_{S}(^3 {\rm He}) \approx f_4(^3 {\rm He}) \tau_{S}(^4 {\rm He})$.  The recoil kinetics and the Doppler width are both 
modified by the mass difference between \Hethree\ and \Hefour.  To include \Hethree\ in the Monte Carlo, we add scattering functions that account for 
the optical depth, mass difference and isotope shift in \Hethree.  Photons are injected to the Monte Carlo in either the \Hethree\ or \Hefour\ lines, 
i.e. their initial frequency distribution is the sum of two Voigt profiles with normalizations in the ratio $f_4(^3 {\rm He})$.  A photon is 
considered to have escaped the line when it either redshifts out or photoionizes \HI.

We track a combined \Hefour/\Hethree\ occupation fraction and escape probability for both isotopes in the level code.  In principle,
one should use the Monte Carlo to develop a $2 \times 2$ matrix of probabilities for absorption of photons by \Hethree/\Hefour\ given 
that it was emitted in an incoherent process in \Hethree/\Hefour, and then develop individual rate equations for \Hethree\ and \Hefour\ 
occupations in the level code.  We expect that this would be a small correction because including \Hethree\ in the first place has a negligible effect; 
however this assumption should be revisited in future work.

\begin{figure}
\includegraphics[width=3.2in]{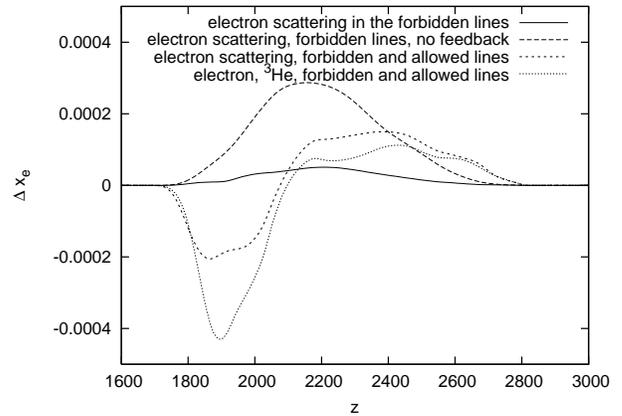}
\caption{\label{figs:e3he_history}  Comparing the effect of Thomson scattering, \Hethree\ scattering, and feedback in the forbidden and allowed lines.  
The uppermost curve is the difference between two models with and without Thomson scattering in the \InterHe\ and \QuadHe\, where neither has feedback.  
This retards recombination because, on average, more photons are injected into the optically thick region of the line.    (Note that once feedback of the 
radiative distortion is added, the effect of Thomson scattering is greatly reduced.)  Thomson scattering in \LyaHe\ decreases the escape rate at early times, 
but once \HI\ opacity becomes significant, more photons are on average removed from the optically thick part of the line.  \Hethree\ can be seen to accelerate 
recombination slightly by assisting photons out of optically thick regions at late times.  We note, though, that the typical error induced by resampling the Monte 
Carlo is of order $10^{-4}$ (see Fig.~\ref{figs:convergence}).  To confirm the effect of \Hethree\ would take a significantly finer grid of probabilities, but the 
actual value is immaterial to \HeI\ recombination overall, if it is this small.}
\end{figure}

Fig.~\ref{figs:e3he_history} shows the cumulative effect of Thomson scattering of Sec.~\ref{sec:thom} and \Hethree\ scattering.  The variation in $x_e$ 
from \Hethree\ scattering is negligible and of order $2 \times 10^{-4}$, which is of the same order as the error induced by resampling the escape 
probability grid -- indeed, to confirm that there {\em is} any effect at all would require significantly more Monte Carlo information.

\section{Additional effects and sources of error}
\label{sec:additionalsys}

\subsection{Rare decays}
\label{ss:rare}

In Paper I we considered the effect of feedback and continuous opacity in the \LyHe, \InterHe, \QuadHe\ transitions in \Hefour.  The spontaneous 
transition rates for \QuadHe\ \cite{2002JPhB...35..421C} and the allowed \cite{1984PhRvA..29.2981K, 1988ApJ...329..493K, 1993A&A...275L...5C, 
1989ApJ...336..504K, 1949RSPTA.242..101B} lines are well-known, but there is considerable variation in the rates described in intercombination 
literature.  Three accurate methods of finding the intercombination rate exist, namely the relativistic random phase approximation 
\cite{1977PhRvA..15..154L}, Hylleraas-type \cite{1930ZPhy...65..209H} methods \cite{1969ApJ...157..459D}, and the $Z$-expansion 
\cite{1978JPhB...11L.391L}.  We used the rates from Ref.~\cite{1978JPhB...11L.391L} as these were available for the greatest number of states.  

We have not resolved the fine structure $J$-levels in ortho-\HeI\ in our multilevel atom code, assuming instead that each level is populated according 
to its statistical ratios.  This assumption is valid because the only rates that depend on $J$ (instead of merely the quantum numbers $n$, $S$, and 
$L$) are the intercombination line rates, and these are all slow compared to allowed transitions in ortho-\HeI\ that mix different values of $J$.  For 
example, mixing among the $2^3P^o_{0,1,2}$ levels by emission followed by absorption of a 1.08$\,\mu$m photon is $\sim 10^5$ times faster than the 
intercombination decay $2^3P^o_1\rightarrow 1^1S$.  The fine structure also does not affect the Voigt profile of the \HeI] \InteraHe\ line because, on 
account of angular momentum conservation (which is exact) only the $J=1$ fine structure level can participate in an electric dipole line connecting to 
$1^1S$.  In the \HeI] \InteraHe\ literature, the quoted spontaneous decay rate \cite{1978JPhB...11L.391L} is for $J=1$, which accounts for only 3 of 
the 9 possible states in the $2^3P^o$ terms that are not resolved in the level code.  Thus, the spontaneous rate used here, $A_{2^3P^o\rightarrow 1^1S}$ 
is $1/3$ of the typical literature values.

\begin{figure}
\includegraphics[width=3.2in]{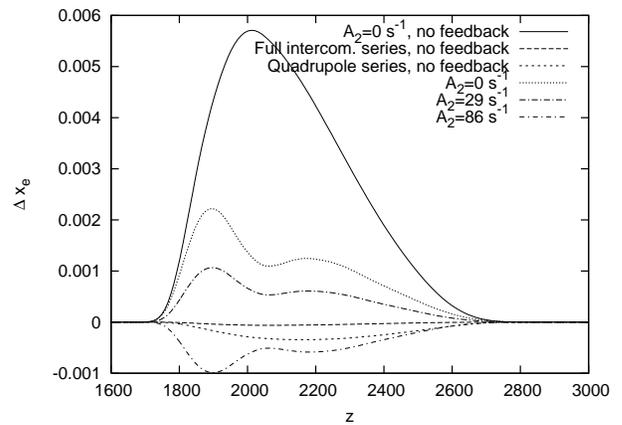}
\caption{\label{figs:intercombinationeffect} There is significant disagreement in the literature describing the intercombination rate because of the 
approximations involved.  Here we show that the influence of a $\pm 50\%$ difference in the \InteraHe\ spontaneous decay rate (here denoted $A_2$) 
on $\Delta x_e$ is roughly $<1 \times 10^{-3}$ relative to the rate used throughout of $57~{\rm s}^{-1}$.  The uppermost plot shows the case where 
the \InteraHe\ rate is neglected entirely.  Once the feedback of non-thermal 
distortions between lines is taken into account, the overall effect of the intercombination lines becomes much less significant.  (Note that all the 
escape probabilities used here include continuous opacity, developed in Paper I).  Also included are the cases where the entire \InterHe\ series is 
included, and the \QuadHe\ series are included in the base model.  The $n>2$ intercombination rates are truly negligible, and the contribution from 
\QuadHe\ is of order, or less than the uncertainty due to the \InteraHe\ rate.}
\end{figure}

Fig.~\ref{figs:intercombinationeffect} considers the effect of variation in the \InteraHe\ spontaneous rate, the contribution of intercombination transitions from 
$n>2$ (which are taken from \cite{1978JPhB...11L.391L} and scaled by $n^{-3}$ for higher levels), and the \QuadHe\ series.  These uncertainties are 
shown to be considerable for \HeI\ recombination, and outweigh many more subtle effects.  (Also note that the magnetic-quadrupole transition \MquadHe\ 
has a spontaneous rate $\sim 0.4~{\rm s}^{-1}$ \cite{1969ApJ...158.1199D}, and is negligible.) 

There is also a metastable triplet level $2^3S_1$ with a forbidden one-photon decay, but its rate is very small, $A=1.27\times 10^{-4}\,$s$^{-1}$ 
\cite{1971PhRvA...3..908D}. This line has not been included because it does not contribute significantly to the formation of ground-state \HeI.  A 
rough estimate of its importance relative to decays from the metastable singlet level can be obtained as follows.  The ratio of level populations will 
be roughly $3e^{\Delta E/\kB T_{\rm r}}$ (where $\Delta E/\kB=9240\,$K) since the excited levels are in equilibrium during \HeI\ recombination to a 
very good approximation.  The ratio of rates of formation of ground-state \HeI\ via decay from $2^3S$ versus $2^1S$ will then be $\sim 3e^{\Delta E/\kB 
T_{\rm r}}(A_{\rm triplet}/\Lambda_{\rm singlet})$, which is $<2.2\times 10^{-4}$ at $z>1000$, which is negligible.

Another route to the ground state is the electric octupole decay [\HeI] $n^1F^o$--$1^1S$ with $n\ge 4$.  The rates for these have not been calculated 
to our knowledge, however generically one expects them to be suppressed relative to the electric quadrupole transitions by a factor of ${\cal 
O}(\alpha^2)$.  Given that the quadrupole decays produced a change in the ionization history of order $|\Delta x_e|_{\rm max}\sim 3\times 10^{-4}$, and 
given that the octupole decays come from higher-energy levels than $3^1D$ with correspondingly lower abundances, we expect a smaller effect from the 
octupole series (the effect need not be a factor of $\sim\alpha^2$ less because the [\HeI] $3^1D$--$1^1S$ line is optically thick during much of helium 
recombination).  In fact, the effect of including octupole transitions is probably much less than quadrupole transitions because they overlap: the 
$4^1F^o$--$1^1S$ and $4^1D$--$1^1S$ lines are split by $\Delta\nu/\nu=2.8\times 10^{-5}$, which is 2.5 Doppler widths at $z=2000$, and the splitting is 
less for higher $n$.

\subsection{Collisional effects in \HeI} 
\label{ss:collisionsmain}

Collisions of atoms and ions with electrons play a key role in determining level populations in the interstellar medium.  In general, they are less 
prominent during cosmological recombination because of the very high density of photons.  There are four major collisional processes that affect the 
recombination history: (1bb) re-distribution of excited levels of \HI, \HeI, or \HeII; (1bf) collisional ionization or its inverse, three-body recombination;
(2) excitation/de-excitation of an atom from/to the ground state; and (3) charge transfer reactions.  We argue that these considerations can be 
neglected for \HeI\ recombination.  

Here the discussion is limited to collisional processes in \HeI\ during \HeI\ recombination.  Because collisional processes generally draw species into 
equilibrium, and \HI\ is very nearly in equilibrium during \HeI\ recombination, collisional processes in \HI\ can be neglected here. Moving in to \HI\ 
recombination, especially in its late phases, \HI\ collisional processes are extremely important as the levels fall out of equilibrium and $l$ and 
$l'$ sublevels can be easily shuffled by collisional processes \cite{2007MNRAS.374.1310C}.  Late time effects in \HI\ recombination deserve much 
attention, but are beyond the scope of this paper.

\subsubsection{Equilibrating $2^1S$ by rates connecting to $n>2$}

\begin{figure}
\includegraphics[width=3.2in]{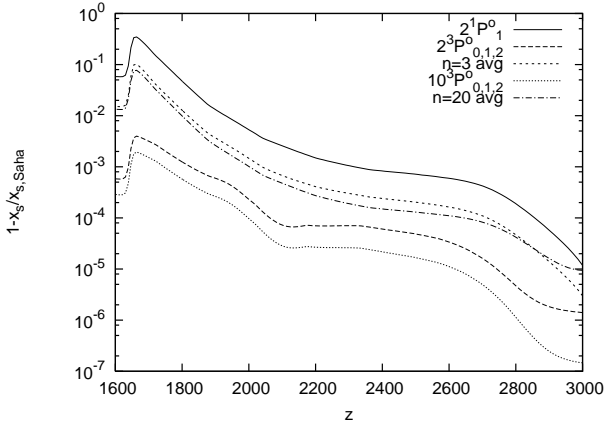}
\caption{\label{figs:leveldeparture} The departure of several levels from equilibrium values during \HeI\ recombination.  The excited states begin to 
fall out of equilibrium for $z<2000$.  Significant among these are $2^1P^o$ and $2^1S$, departing from Saha values by of order $10\%$ and larger by $z=1700$. 
(Note that the departure histories of $2^1P^o$ and $2^1S$ are so similar that here we only show $2^1P^o$.  Likewise with $2^3S$ and $2^3P^o$, so only 
$2^3P^o$ is shown.)  A significant concern is that collisional rates could drive $2^1P^o$ and $2^1S$ closer to equilibrium and modify the overall $1^1S$ 
formation rate.}
\end{figure}

During \HeI\ recombination, the highly-excited populations in both \HeI\ and \HI\ are very close to Saha-Boltzmann equilibrium.  Fundamentally, 
rates for electron collisions inducing bound-bound and bound-free transitions are small then because they depend on departure from equilibrium.  
Fig.~\ref{figs:leveldeparture} compares the departure of several highly-excited states from their Saha occupations (here, without collisional effects
added, which would push the levels further into equilibrium).  We note that $2^1P^o$ and $2^1S$ depart from 
their Saha occupations by tens of percent at $z \sim 1700$.  Collisional processes will tend to push these closer to equilibrium with the continuum,
and this will modify the formation rate to $1^1S$.  We will see that the collisional rates connecting $2^1P^o$ to excited levels are not sufficient to 
affect its departure from Saha.
 
The electron-collision induced transition rates are given by the standard rate equations which respect detailed balance
\beq \dot x_n \biggl |_{coll,bf} = \langle \sigma v \rangle_{cn} D_{ci} n_e, \eeq
where
\beq D_{ci} = g_n e^{-E_n/(kT_m)} \left ( \frac{ h^2 }{ 2 \pi m_e k T_m } \right )^{3/2} \frac{ x_e}{ 2} - x_n \eeq
for collisional ionization and three-body recombination.
For the collisional bound-bound rates, we have
\begin{eqnarray}
\dot x_n \biggl |_{coll,bb} &=& \sum_{n'>n}  \langle \sigma v \rangle_{n' \rightarrow n} D_{n,n'} n_e \\
&&- \sum_{n'<n}  \langle \sigma v \rangle_{n \rightarrow n'} D_{n',n} n_e,
\label{eqn:collbbrates}
\end{eqnarray}
where
\beq D_{n,n'} = x_{n'} - x_n \frac{g_{n'} }{ g_{n}} e^{(E_n - E_{n'})/(kT_m)}. \eeq
(Note that between the excitation and de-excitation we have reversed the subscript order $D_{n',n}$ to $D_{n,n'}$, respectively.)
The total rate $n' \rightarrow n$ to a level $n$ then depends on the population of $n'$, the departure of $n$ from Saha equilibrium and 
the process's rate coefficient, here $\langle \sigma v \rangle n_e$, as in radiative bound-bound processes.  We can thus compare the collisional rates
to the spontaneous rate $A_{n' \rightarrow n} P_S$ connecting the levels.  For the allowed transitions $2^1S-n^1P^o$, 
$\langle \sigma v \rangle$ is given by Ref.~\cite{2000A&AS..146..481B}:
$\langle\sigma v\rangle\sim \{ 5 \times 10^{-7}, 1 \times 10^{-8}, 4 \times 10^{-9},  2 \times 10^{-9} \} T_4^{-1/2} \,$cm$^3\,$s$^{-1}$ for $n=2,3,4,5$ respectively.
These are all between 9 and 12 orders of magnitude smaller than the radiative Einstein coefficients connecting these levels (for $n_e = 1080~{\rm cm}^{-3}$). 
The radiative rates connecting levels are not sufficiently suppressed by their escape probabilities for collisional effect to contribute (recall the allowed \LyaHe\ 
series is optically thick, but transitions from and to $n\ge2$ are thin to a good approximation).  

To assess whether collisional processes through radiatively forbidden lines can 
push $2^1S$ closer to equilibrium, we can compare the total collisional rate into $2^1S$ to the total radiative bound-free rate supplying $2^1S$.  If 
we consider decays from from $n'$ to $n$ in Eq.~(\ref{eqn:collbbrates}), and define the departure from equilibrium as $b_n = x_n/n_{n,{\rm Saha}}$, the 
sum becomes
\begin{eqnarray} \dot x_n |_{n' \rightarrow n} &=& x_{n, Saha} (1-b_n) \nonumber \\
&& \times \sum_{n'>n} \biggl ( \frac{g_n'}{g_n} e^{-h\nu_{n,n'}/(k_B T_m)} \nonumber \\
&& \times \langle \sigma v \rangle_{n' \rightarrow n} n_e \biggr ), \end{eqnarray}
while for thermal radiation, the rate is
\beq \dot x_n |_{bf} = x_{n,{\rm Saha}} (1-b_n) \beta_{2^1S} . \eeq
When $2^1S$ begins to fall out of equilibrium the typical collisional rate coefficient $\langle \sigma v \rangle n_e$ to all other levels \cite{2000A&AS..146..481B}  is between 
$10^{-5}~{\rm s}^{-1}$ and $10^{-7}~{\rm s}^{-1}$.  Combined with the factor exponentially suppressing transitions of increasing energy separation, it is clear that $n=2$ 
dominates the sum.  Yet, these rates are dwarfed by the radiative bound-free rates to the level, for which $\beta_{2^1S}  \sim 10^{3}~{\rm s}^{-1}$ around $z \sim 1700$.

\subsubsection{De-excitatation of the $n=2$ family to $1^1S$} 

It is possible to speed up recombination if collisions efficiently de-excite helium atoms to their ground state without producing further ionizing 
photons.  In the case of \HeI\ there is the possibility that the collisional de-excitation of the metastable $2^3S$ level could compete with two-photon decay, 
which starts from the higher-energy $2^1S$ level.  Assuming approximate equilibrium between $2^1S$ and $2^3S$, $n(2^3S)/n(2^1S)\approx 
3e^{\Delta E/kT_r}\approx 25$ at $z \sim 1600$.  The collision strengths of Ref.~\cite{1990JPhB...23.4321S,2000A&AS..146..481B} give $\langle\sigma v\rangle\sim 2\times 
10^{-9}T_4^{-1/2}\,$cm$^3\,$s$^{-1}$ where $T_4$ is the temperature in units of $10^4\,$K.  The ratio of collisional de-excitation to two-photon 
de-excitation is then \beq \frac{n(2^3S)}{n(2^1S)}\frac{n_e\langle\sigma v\rangle_{2^3S\rightarrow 1^1S}}{\Lambda_{\rm HeI}}, \eeq which is of order 
$10^{-6}$ throughout \HeI\ recombination and hence can be neglected.  Considering the rest of the $n=2$ family, 
$\langle\sigma v\rangle\sim \{ 0.2, 1, 0.5 \} \times 10^{-9} T_4^{-1/2} \,$cm$^3\,$s$^{-1}$ for decays $2^1P^o\rightarrow 1^1S$, $2^1S \rightarrow 1^1S$ 
and $2^3P^o \rightarrow 1^1S$, respectively.  As in the case of $2^3S$, the relative occupations in the $n=2$ family are not sufficiently different to 
enhance overall collisional de-excitation rates for these processes to matter.  Similar collision strengths for de-excitation to the ground state are obtained 
from higher levels of \HeI, but their occupation probabilities are much smaller, hence they will contribute even less to the formation of \HeI\ $1^1S$.

For \HeII\ recombination, the radiative rates are sufficient to maintain Saha equilibrium at all times to an accuracy of $0.2 \%$ in $x_e$ (much less in $C_\ell$),
and this circumstance cannot be altered by including collisions.
 
 \subsubsection{Charge transfer}

Charge transfer reactions will accelerate \HeI\ recombination and proceed through
\begin{equation}
{\rm He}^+ + {\rm H} \leftrightarrow {\rm He} + {\rm H}^+ + \gamma.
\end{equation}
(The nonradiative reaction is much slower due to the large separation between potential energy surfaces for the initial and final states 
\cite{1989PhRvA..40.2340Z}.)  This is similar to continuous opacity, examined in Paper I, where a \HI\ population can assist \HeI\ recombination.  
(Indeed, because continuous opacity acting within the line accelerates recombination as soon as a small \HI\ population is present, \HI\ populations are not sufficient 
during \HeI\ recombination for charge transfer to occur.)  The total charge transfer contribution to the rate is given by

\beqa
\dot x_{\rm HeI} &=& (1.2 \times 10^{-15} {\rm cm}^3~{\rm s}^{-1}) \left (\frac{T_{\rm m}}{300~\rm K} \right )^{0.25} n_{\rm H} \nonumber \\
&&\times \left[
x_{\rm HI}  x_{\rm HeII} - \frac{1}{4}  x_{\rm HII}  x_{\rm HeI}  e^{-\Delta\chi/\kB T_{\rm m}} \right],
\eeqa
where $\Delta\chi=\chi_{\rm HeI} - \chi_{\rm HI}$ is the ionization potential difference.  Note that the second (detailed balance) term is only correct 
if the radiation and matter temperatures are the same, because it involves a photon; however this is the case throughout \HeI\ recombination.
The rate coefficient is given by the fitting formula of Ref.~\cite{1998ApJ...509....1S}.  The charge transfer rate reaches a maximum at $z \sim 
1900$ (if we turn it on in the interval $z\sim1700$ to $z\sim 2200$), where $|dx_{\rm HeI}/d\ln(1+z)| < 2\times 10^{-5}$.  Thus it can be neglected.

\subsection{Peculiar velocities}
\label{ss:peculiar}

It has been suggested in the past that peculiar velocities could be important for the recombination history of the universe.  Taken at face value they
could substantially affect the line profiles: typical baryonic peculiar velocities before matter-radiation decoupling are of order
several$\times 10^{-5}c$ as a
consequence of the primordial fluctuation level, as compared with typical thermal velocities $\sim (k_{\rm B}T_{\rm m}/m_{\rm He})^{1/2}\sim 10^{-5}c$.  The
similarity is (to our knowledge) a coincidence, but it implies that peculiar velocities could lead to significant broadening of the line profile.
Seager et~al. \cite{2000ApJS..128..407S} neglected this possibility because they treated line transfer using the Sobolev escape probability, which does
not depend on the line profile so long as it is narrow.  In contrast, our calculation finds a very significant influence of the line profile on the
recombination history, so the issue of line broadening from peculiar velocities must be considered more carefully.  We argue here that there is in fact
{\em not} a major effect, at least in the standard cosmological model, because the peculiar velocities are coherent on large scales and hence the
peculiar velocity gradients are small.  Our argument may not apply to small-scale peculiar velocities generated in more exotic scenarios
\cite{1998ApJ...503...67W, 1998MNRAS.297.1245S}, and it is even conceivable that recombination may provide a way of constraining these.  Consideration
of such exotic scenarios is beyond the scope of this paper.

In some astrophysical applications, it is possible to treat peculiar velocities
as providing an extra, non-thermal increase in the line width.  This
``microturbulence'' limit applies when the mean free path is large compared to
the coherence scale of the velocity field, and in this case it is only the
distribution of velocities that matters (or only the rms velocity, if the
velocity is Gaussian).  The opposite limit is where the peculiar velocities
have a long wavelength relative to the distance a photon can travel over the
relevant timescale.  In this case, the velocity gradient tensor
$\partial_iv_j$ is constant over the region sampled by a photon, and it is
instead the velocity gradient (in combination with other scales in the problem)
that matters.  One can see the difference between these two cases by considering
what happens to a photon's frequency in the baryon reference frame: in the
microturbulence case the frequency oscillates wildly around some mean value as
the photon passes through eddies of different baryon velocity, whereas in the
uniform gradient case the photon's frequency decreases linearly, with the
peculiar velocity gradient providing a correction to the Hubble constant (and
providing it with direction dependence since $\partial_iv_j$ will in general
have a spin-2 component).  Since most of the velocity power during recombination
is on scales of several Mpc comoving (the Silk damping length; see below),
whereas the Thomson mean free path $(an_{\rm e}\sigma_{\rm T})^{-1}$ is hundreds
of kpc comoving (and even less for photons near resonance lines), microturbulence
does not apply here and a more sophisticated analysis is necessary.  We
investigate the nature of the baryon velocity field in Sec.~\ref{sss:v1}, and
then consider how far a photon can travel during the time it spends in a
resonance line in Sec.~\ref{sss:v2}.  Most of the numbers describing the velocity
field and photon transport here will be quoted at $z=2000$, however the
qualitative picture is the same at other redshifts during helium recombination.

\subsubsection{The baryon velocity field}
\label{sss:v1}

\begin{figure}
\includegraphics[angle=-90,width=3.2in]{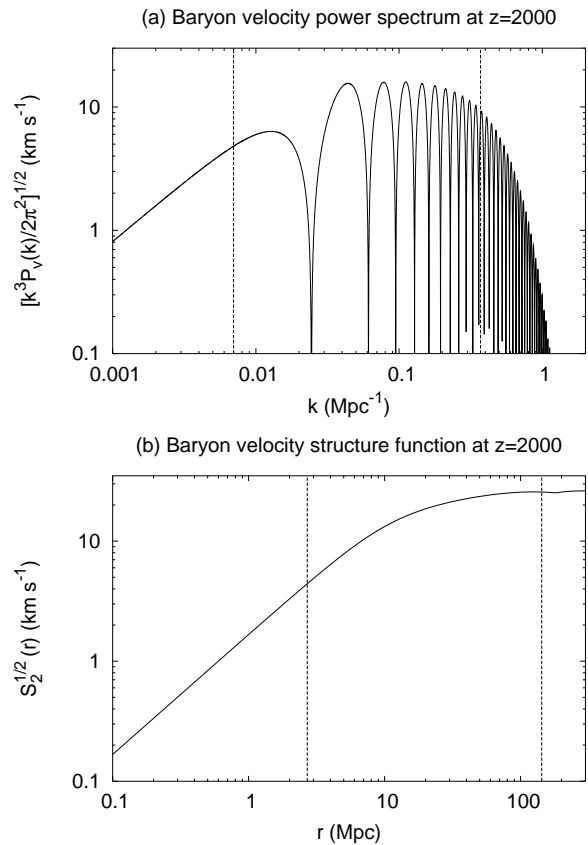}
\caption{\label{fig:vpec}The peculiar velocity (a) power spectrum and (b) structure
function of the baryons at $z=2000$.  The vertical lines show the Silk damping scale
($k_{\rm D}=0.37\,$Mpc$^{-1}$, or $k_{\rm D}^{-1}=2.7\,$Mpc in the lower panel) and
the Hubble scale $aH/c=0.007\,$Mpc$^{-1}$ or $c/aH=143\,$Mpc.}
\end{figure}

We compute the baryon velocity power spectrum $P_v(k)$ using the Boltzmann code
{\sc cosmics} \cite{1995ApJ...455....7M} in Newtonian gauge.  We assumed a
primordial power spectrum with $\Delta^2_{\cal R}=2.4\times 10^{-9}$ at
$k=0.002$ and $n_s=0.95$, consistent with the recent WMAP results
\cite{2007ApJS..170..377S}.  We then obtain the velocity structure function at
comoving separation $r$,
\begin{equation}
S_2(r) \equiv \langle [{\bf v}({\bf 0})-{\bf v}({\bf r})]^2\rangle
= \int 2[1-j_0(kr)]\frac{k^3P_v(k)}{2\pi^2}\,\frac{dk}k,
\label{eq:structure}
\end{equation}
where $j_0$ is the spherical Bessel function.  The result is shown at $z=2000$
in Fig.~\ref{fig:vpec}.  Note the importance of the Silk damping scale
$k_{\rm D}^{-1}=2.7\,$Mpc (as computed using the formula in
Zaldarriaga \& Harari \cite{1995PhRvD..52.3276Z}): perturbations at
$k\gg k_{\rm D}$ are damped out, with the result that at separations
$r\ll k_{\rm D}^{-1}$ the relative velocities of packets of baryon fluid are
dominated by expansion and shearing due to long-wavelength perturbations
($k^{-1}\gg r$).  In this limit we find
\begin{equation}
S_2(r) \approx \frac13(\theta_{\rm rms}ar)^2,
\label{eq:s2}
\end{equation}
where
\begin{equation}
\theta_{\rm rms}^2 = \int \left(\frac ka\right)^2\frac{k^3P_v(k)}{2\pi^2}\,\frac{dk}k
\end{equation}
is the rms peculiar expansion of the baryon fluid.
At $z=2000$ this is $\theta_{\rm rms}a=2.9\,$km$\,$s$^{-1}\,$Mpc$^{-1}$.
(In principle the small-scale structure function depends on the shear as well
as the expansion, however for scalar perturbations they are not independent.)
On scales small compared to $k_{\rm D}^{-1}$ the baryon velocity field essentially
has a uniform gradient $\partial_iv_j$ (though its expansion rate may be perturbed
and its shear nonzero); on larger scales the Universe contains many patches of
different baryon velocity.

The velocity field at other redshifts of interest ($1700<z<3000$) is qualitatively
similar, the major differences being that the Silk scale is shorter at higher
redshift, and the acoustic scale (which sets the oscillations in the power spectrum)
is also shorter at high $z$.  The rms peculiar expansion declines from
$\theta_{\rm rms}a=5.4\,$km$\,$s$^{-1}\,$Mpc$^{-1}$ at $z=3000$ to
$\theta_{\rm rms}a=2.2\,$km$\,$s$^{-1}\,$Mpc$^{-1}$ at $z=1700$.  It only undergoes
qualitative changes after kinetic decoupling ($z\sim 1100$) after which the baryons
fall into the dark matter potential wells.  This later era is not relevant to this
paper, although it could affect the late stages of hydrogen recombination.

Before we continue, it is worth noting that if Silk damping were somehow turned off,
then the velocity fluctuations of order $10\,$km$\,$s$^{-1}$ seen in
Fig.~\ref{fig:vpec}(a) would continue out to arbitrarily large $k$.  These
small-scale fluctuations would be in the microturbulence regime, and because they
are larger than the helium thermal velocity, they would significantly broaden the \HeI\ lines.
In Appendix~\ref{app:cpt}, we describe cosmological perturbation theory relevant for radiative
transport in helium in more detail, considering coupling of helium to the other fluids,
damping, and plasma oscillations.

\subsubsection{Photon transport}
\label{sss:v2}

We now ask how far a photon can diffuse during the time it spends in a resonance line.
By combining this information with the results from Sec.~\ref{sss:v1}, we will be able
to determine the typical velocity differences within this region, and what effect (if any)
they will have on recombination.  We consider the two lines most relevant for helium
recombination -- the allowed \HeI\ \LyaHe\ line, in which a photon typically undergoes
many scatterings during passage through the line, and the semiforbidden
\HeI] \InteraHe\ line, which has optical depth of at most a few.

For the allowed \HeI\ \LyaHe\ line, we wish to know how far the photon can travel as
it redshifts from
$\nu_+=\nu_{\rm line}+\Delta\nu_{\rm line}$ to $\nu_-=\nu_{\rm line}-\Delta\nu_{\rm line}$.
(Recall that $\Delta\nu_{\rm line}$ is the detuning beyond which the integrated optical
depth to incoherent processes is unity.  We only included incoherent processes in this
definition since coherent scattering does not change the ionization/electronic state of
the atom.)  Since we have $f_{\rm inc}\ll 1$, the coherent scattering optical depth is
very large even in the incoherent-thin part of the damping wings.  Therefore the photon
moves by diffusion, and we should calculate its rms diffusion distance,
\begin{eqnarray}
L_{\rm diff} &=& \frac1a\left[ \int cL_{\rm mfp}\,dt \right]^{1/2}
\nonumber \\
&=& \frac1a\left[ \int c\frac c{H\nu_{\rm line}\,d\tau/d\nu}\,\frac{d\nu}{H\nu_{\rm line}} \right]^{1/2}
\nonumber \\
&=& \frac c{aH\nu_{\rm line}}\left[ \int_{\nu_-}^{\nu_+} \frac{d\nu}{\tau_{\rm S}\phi(\nu)} \right]^{1/2}.
\end{eqnarray}
(Here $L_{\rm mfp}$ is the physical mean free path.  The denominator in the last integral
is $d\tau/d\nu$; note that this is an upper limit if Thomson opacity is also important.)
The integral is dominated by the damping wings where $\phi(\nu)=\Gamma_{\rm line}/4\pi^2\Delta\nu^2$, so we have
\begin{eqnarray}
L_{\rm diff} &=& \frac{2\sqrt 2\,\pi}{\sqrt 3} \frac {c\Delta\nu_{\rm line}^{3/2}}{aH\nu_{\rm line}\tau_{\rm S}^{1/2}\Gamma_{\rm line}^{1/2}}
\nonumber \\
&=& \frac 1{2\sqrt 6\,\pi^2} \frac {c\Gamma_{\rm line}\tau_{\rm S}f_{\rm inc}^{3/2}}{aH\nu_{\rm line}},
\end{eqnarray}
where in the last line we have substituted
$\Delta\nu_{\rm line}=\Gamma_{\rm line}\tau_{\rm S}f_{\rm inc}/4\pi^2$ (see Paper I).
At $z=2000$, we have $\tau_{\rm S}=2.0\times 10^7$, $f_{\rm inc}=2.8\times 10^{-3}$, giving
$L_{\rm diff}=3.1\,$kpc.

For the semiforbidden line, the damping wings are negligible and the profile is Gaussian
with width $\Delta\nu_{\rm D}$.  For most of the recombination history we cannot define
an optically thick part of the line in analogy to $\Delta\nu_{\rm line}$, because the
optical depth is $\tau_{\rm S}<2$ and hence either half of the line is optically thin.
We can however calculate the comoving straight-line distance a photon must travel before
it redshifts by $\Delta\nu_{\rm D}$, which (since the optical depth never exceeds a few)
should serve as a guide to the typical distance a photon can travel while interacting with
the semiforbidden line.  This distance is
\begin{equation}
L_{\rm sl} = \frac{c\Delta\nu_{\rm D}}{aH\nu_{\rm line}},
\end{equation}
or 2.2 kpc at $z=2000$.  This is an upper limit on the net distance traveled by the
photon because the latter could be reduced by scattering (although for optical depths of
a few this reduction is likely to be modest).  Similar results hold at other redshifts:
at $1700<z<3000$, we find that $L_{\rm sl}$ has a peak value of 2.3 kpc, and
$L_{\rm diff}$ has a peak of 4.0 kpc.

It is easily seen that the distances $L_{\rm diff}$ and $L_{\rm sl}$ are much less than
the Silk length.  This is not surprising: the Silk length is (roughly) the distance a
photon can diffuse during a Hubble time, whereas we are asking how far a photon can go
during the much shorter time required to redshift through a line.  These distances
correspond to scales over which the rms velocity difference is
$S_2^{1/2}(L_{\rm diff})=9\,$m$\,$s$^{-1}$ and $S_2^{1/2}(L_{\rm sl})=6\,$m$\,$s$^{-1}$.
In comparison, the rms thermal velocity per coordinate axis is
$(k_{\rm B}T_{\rm m}/m_{\rm He})^{1/2}=3.4\,$km$\,$s$^{-1}$.  Thus over the region that
a photon explores during its transport through a line, the peculiar velocity differences
are very small compared to the thermal velocities of the helium atoms.  This is true even
though the baryons have a bulk ($>$Mpc scale) flow at an rms velocity of $18\,$km$\,$s$^{-1}$.

Even though the relative peculiar velocities are small compared to the thermal velocity,
it is still possible that the velocity gradient can lead to a net effect on the photon's
frequency after many scatterings.  In order to assess this possibility, we measure the
photon's frequency in the baryon frame rather than in the Newtonian frame, as this leaves
all scattering terms in the radiative transfer equations unchanged.  It also eliminates
the explicit dependence of position on the problem, so long as the velocity gradient is
uniform over the region explored by the photons (which is true since
$L_{\rm sl},L_{\rm diff}\ll k_{\rm D}^{-1}$).  In this case, the photon is moving through
a medium whose expansion rate (i.e. velocity gradient) is not exactly given by the Hubble
rate $H$, but rather by a correction $H+\theta_b/3$, where $\theta_b$ is the expansion
perturbation.  The velocity gradient may also have a shear, which for scalar perturbations
is generally of the same order as $\theta_b$.  The Hubble expansion term
$\dot\nu|_{\rm Hubble}=-H\nu$ in the radiative transfer equation is replaced:
\begin{equation}
-H\nu \;\rightarrow\;
-\left(H + \frac13\theta_b + \sigma_{b,ij}n^in^j\right)\nu,
\label{eq:hreplace}
\end{equation}
where $\sigma_b$ is the baryon shear tensor and $n^i$ is the direction of propagation.
The peculiar expansion has variance $\langle\theta_b^2\rangle = \theta_{\rm rms}^2$.
Since we have for each scalar Fourier mode
$\sigma_{b,ij}({\bf k}) = (\hat k_i\hat k_j-\delta_{ij}/3)\theta_b({\bf k})$,
it follows that $\langle \sigma_{b,ij}\sigma_{b,ij}\rangle = (2/3)\theta_{\rm rms}^2$.

The effect of the $\theta_b$ term in Eq.~(\ref{eq:hreplace}) is equivalent to a local
modification of the Hubble rate, with rms fractional change
$\theta_{\rm rms}/3H=5\times 10^{-4}$.  Its effect can therefore be modeled by changing
the Hubble rate and asking what effect this has on the recombination rate $\dot x_{\rm HeI}$.
The effect of the Hubble rate is to change the escape probabilities, whose rms change will be at most $5\times 10^{-4}$
(escape probabilities scale as $P_{\rm S}\propto H$ for optically thick lines with no continuum
opacity, and the scaling is shallower if there is continuum opacity or if $\tau_{\rm S}$ is of order unity or less).
Thus the typical correction to the recombination rate $\delta\dot x_{\rm HeI}/\dot x_{\rm HeI}$ will be of order $5\times 10^{-4}$.
This correction
will be positive in some regions and negative in others, because $\langle\theta_b\rangle=0$.
Because $\sigma_b$ has spin 2, the change in the recombination rate $\delta\dot x_{\rm HeI}$
can only exist at order $\sigma^2$, and will probably be much smaller than the effect of the
$\theta_b$ term.

\subsection{Additional considerations}
\label{ss:additionalconsiderations}

There are several issues that we have neglected throughout.  The first is the modification to the vacuum dispersion relation from multiple resonant scattering, 
such as through \LyaHe\ \cite{peeblesconv}.  (This is associated with a modification to the refractive index, or alternately the complex dielectric constant and group velocity, in the 
neighborhood of the line.)  Throughout, we have assumed $\omega = c k$ and that the radiation fields are completely described by $\pha(\omega)$ in calculation of the matrix 
elements.  Now both of these assumptions break down, and the issue arises that the spontaneous and stimulated decay rates are modified by the mild dielectric behavior of the gas 
near the resonance frequency.  These rates are the subject of active research 
\cite{1996PhR...270..143L, 1999PhRvA..60.4094S, 1992PhRvL..68.3698B, 1995JMOp...42.1991M, 2006JPhB...39S.627J}.  
Generally, the modifications are directly related to the real part of the refractive index, the group velocity, and the imaginary part of the dielectric constant (from absorption) 
\cite{1995JMOp...42.1991M,1999PhRvA..60.4094S}.  These have maxima $\Re(n) -1 \approx 6 \times 10^{-17}$, $\Im(\epsilon/\epsilon_0) \approx 2 \times 10^{-16}$, $1-v_g \approx 7 \times 10^{-12}$
around the \HeI\ 21.2~eV resonance.  A further treatment would require a significant overhaul of the methods used throughout, and because of 
its subtlety and complexity (how does radiative transport work when the radiation and matter statistics are correlated; what is the appropriate density of states; how can this be 
implemented economically?) would vastly complicate the argument.  We believe that because the pertinent quantities that describe deviations from vacuum propagation 
are so small that this is a negligible addition to the model's physics.  There are lower-frequency plasma effects associated with the plasma frequency (typically several hundred 
kHz here), which leads to negligible dispersion for any of the resonances involved, especially because the relevant quantities depend on $\omega_p^2/\omega_0^2$ for some transition
frequency $\omega_0$.

In paper I, we emphasized the corrections in transport through the \HeI\ line due to a small population of \HI.  The negative hydrogen ion, neutral lithium \cite{peeblesconv}, 
${\rm HeH}^+$, ${\rm H}_2$ and ${\rm H}_2^+$ are also present, albeit with very low occupations, and can act as a sink for \HeI\ resonance photons through photo-detachment/photoionization.  
(negative hydrogen in an important part of the opacity in stellar atmospheres \cite{1988A&A...193..189J})  Taking a cross section for negative hydrogen of 
$\approx 10^{-17}~{\rm cm}^2$ \cite{1988A&A...193..189J} and assuming Saha populations, the differential opacity is $\approx {\rm few} \times 10^{16}$ times lower than \HI\ (near the 
end of \HeI\, when the negative hydrogen depth is greatest), being largely suppressed by the Saha occupation for the binding energy of $0.755$~eV.  While the ionization 
threshold of lithium is $5.39$~eV \cite{2003IAUJD..17E..13M}, its fractional abundance is at the level of $10^{-9}-10^{-10}$ \cite{2001ApJ...552L...1B}.  Using the cross 
sections to \HeI\ \LyaHe\ radiation from \cite{1996ApJ...465..487V}, we find that the differential optical depth to lithium, is a further three orders of magnitude smaller than negative
hydrogen.  Both processes are truly negligible.  

An upper limit to the effect  of H$_2^+$ and HeH$^+$ can be obtained by assuming that their abundance is in thermal equilibrium according to the reactions 
H(H$^+$,$\gamma$)H$_2^+$ and He(H$^+$,$\gamma$)HeH$^+$ (there are other pathways for formation of these species, but they are suppressed here by the abundance of 
the precursors and the temperature), and that their cross sections to UV photons are given by the unitarity bound for electric dipole transitions, $\sigma\le 3\lambda^2/8\pi$.  
We note here that the equilibrium abundance of these ions is rapidly increasing as the universe expands and cools, so the true abundance may be much less than this if the 
reaction rates are slow (see e.g. \cite{2006MNRAS.372.1175H}).  Also photoexcitation and photodissociation cross sections are usually several orders of magnitude less than
the unitarity bound except at the centers of resonance lines, so the analysis here is probably very conservative (i.e. overestimates the effect by a large amount).  The H$_2^+$ 
levels used here are the same as those computed in Ref.~\cite{2006MNRAS.372.1175H} (with a binding energy taken to be $2.65$~eV~\cite{2006MNRAS.372.1175H}), and 
the HeH$^+$ partition function is taken from Ref.~\cite{2005MNRAS.357..471E} (with a binding energy taken to be $1.84$~eV \cite{1998ApJ...508..151Z}.) For H$_2^+$, the 
equlibrium analysis gives $x[$H$_2^+] \sim 4\times 10^{-21}$ at $z=1700$ (the end of \HeI\ recombination), for which the unitarity bound implies a depth of $n\sigma c/H\le 5\times 
10^{-6}$ at 21.2 eV; this number is much less at higher redshift.  For HeH$^+$, we find $x[$HeH$^+] \sim 4 \times 10^{-25}$ at $z=1700$, hence $n\sigma c/H\le 5 \times 10^{-10}$.  
Again this number is much less at higher redshift.  For ${\rm H}_2$, using the partition function from 
\cite{1987A&A...182..348I} (assuming equilibrium populations) we get $x[{\rm H}_2] \sim 2 \times 10^{-23}$ at $z=1700$, giving $n\sigma c/H \le 2.8 \times 10^{-8}$.  Thus 
opacity due to these species will have no significant effect on \HeI\ recombination.

Also neglected is the possibility of a velocity-dependent occupation fraction for the atomic levels.  We have treated the atom's absorption profile conditioned on the struck 
atom's velocity (which is taken to be thermal) in the Monte Carlo developed in Paper I.  We have neglected these effects in incoherent scattering by assuming the photon is 
re-emitted across a pure Voigt profile at line center.  These corrections are expected to be small, but warrant consideration.
 
\subsection{Convergence of the methods}
\label{ss:convergence}

Here we check the convergence and accuracy of the numerical methods employed throughout this series of papers.  The most significant among these for 
the overall rate are the level of refinement of the probability grids estimated in the Monte Carlo and their accuracy (which can be assessed easily by 
resampling) and the accuracy of the atomic level code numerical solution.

\begin{figure}
\includegraphics[width=3.2in]{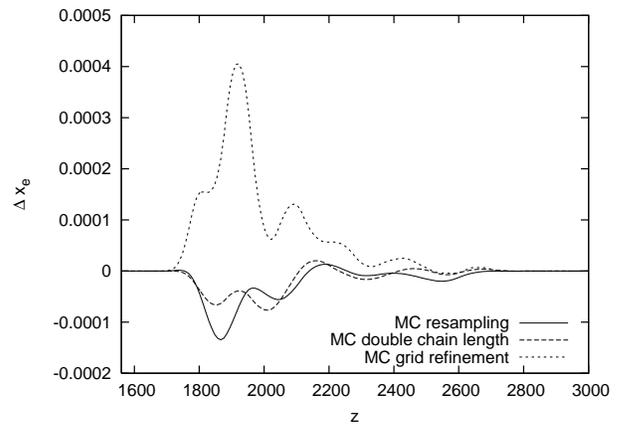}
\caption{\label{figs:convergence} Comparing several numerical convergence issues in the Monte Carlo-estimated $11\times 21$ (redshift by $x_{\rm HeI}$) grid of escape probabilities.  Both doubling number of photons in the sample and resampling the Monte Carlo give corrections of order $< 2 \times 10^{-4}$.  Grid refinement is a more significant systematic, roughly $< 4 \times 10^{-4}$, and indicates that log-log interpolation on the coarser grid over-predicts the escape velocity.  Halving the step size in the level code results in error $| \Delta x_e |  < 3 \times 10^{-5}$.}
\end{figure}

\begin{figure}
\includegraphics[width=3.2in]{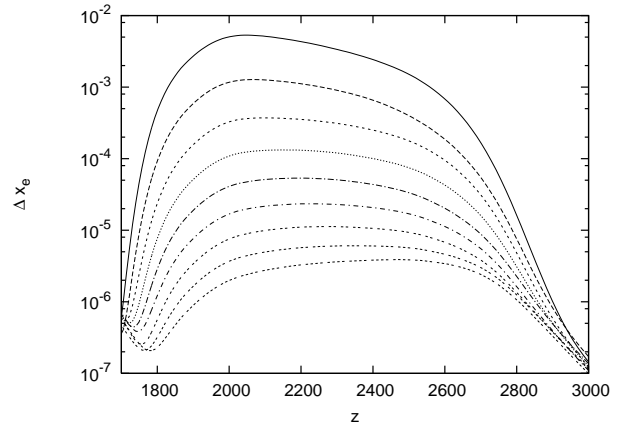}
\caption{\label{figs:feedback_convergence} Convergence of the iterations to include feedback of non-thermal distortions between lines.  These are descending from the difference between no feedback and one interation, between the first iteration and the second, and so on.  Note that by the 4th iteration, the effect is roughly $\Delta x_e < 10^{-4}$, so by going a fifth iteration, any systematic effect is negligible.  Note that the integration tolerance taken in the level code is $1 \times 10^{-5}$.}
\end{figure}

Here we consider three refinement and resampling cases: 1) doubling the number of MC photons in the sample in the $11 \times 21$ (redshift by $x_{\rm 
HeI}$) grid of probabilities, 2) resampling the $11\times 21$ grid, and, 3) refining to a $21\times 41$ grid.  These are shown in in 
Fig.~\ref{figs:convergence}.  We also confirm that the feedback iterations have converged (in the sense of giving negligible differences between 
subsequent iterations) in Fig.~\ref{figs:feedback_convergence}.  We also considered the level code step size.  By doubling the number of steps through 
the recombination history, we change the recombination history by $| \Delta x_e | < 3 \times 10^{-5}$.  

In this series of papers, we have ignored levels with $n>n_{\rm max}$.  As the matter and radiation temperatures drop, $n_{max}$ must increase to 
account for levels that have possibly fallen out of equilibrium.  If the net recombination (capture) rate to these high states becomes significant, 
then truncation at $n_{\rm max}$ will generally contribute less to the formation rate of the neutral species.  Here we simply (roughly) halve $n_{\rm max}$ to 
$n_{\rm max} = 45$ (from $n_{\rm max} = 100$), to find that the change in $x_e(z)$ is negligible and of order $| \Delta x_e | < 4 \times 10^{-5}$.  The contribution to the formation rate 
of the ground state from the decay of these highly excited levels is greatly suppressed by the feedback of the spectral distortions they generate.  Indeed, given
the high optical depth in the \LyHe\ lines, a model neglecting feedback over-estimates the contribution of these highly-excited states to $x_e$ by slightly
over an order of magnitude.

\section{Effect on CMB anisotropy}
\label{sec:CMBaniso}

Our principal motivation for developing a detailed understanding of \HeI\ recombination in this series of papers has been to improve the accuracy of CMB 
models.  This section and the following section are devoted to the magnitude of improvements to the CMB power spectrum $C_\ell$, and to the residual uncertainty from \HeI\ 
recombination.  We will argue here that the residual uncertainty is small compared to the dispersion among some of the recently discussed 
recombination histories.  The residual uncertainty from \HeI\ recombination in the CMB spectra still exceeds cosmic variance (because of the large 
number of modes available) and is comparable to the uncertainty from the precision of modern linear Boltzmann codes \cite{2003PhRvD..68h3507S}. This 
uncertainty is effectively confined to a one-parameter family of $\{C_\ell\}$s parameterized by the Silk damping scale.

The helium recombination history affects CMB anisotropies by changing the electron abundance in the redshift range $1500<z<2700$.  In general the change in CMB power spectrum can be written as
\beq
\frac{\Delta C_\ell^{TT}}{C_\ell^{TT}} = \int F_T(z) \Delta x_e(z)\,\frac{dz}{1+z},
\eeq
where the function $F_T$ is written in terms of the functional derivative,
\beq
F_\ell^T(z) = \frac{1+z}{C_\ell^{TT}}\frac{\delta C_\ell^{TT}}{\delta x_e(z)}.
\label{eq:ftz}
\eeq
That is, an increase of $\Delta x_e=0.01$ for $N$ $e$-folds of expansion results in an $F_\ell^TN\%$ change in the temperature power spectrum, so $F_\ell^T(z)$ can be considered as a weighting function that indicates how much $C_\ell^{TT}$ depends on a given point in the recombination history.  For polarization-sensitive experiments one may also define an analogous function $F_\ell^E(z)$.  Since the temperature-polarization cross-spectrum $C_\ell^{TE}$ crosses zero twice every acoustic oscillation, we choose to define
\beq
F_\ell^X(z) = (1+z)\frac{\delta \rho_\ell^{TE}}{\delta x_e(z)},
\label{eq:fxz}
\eeq
where $\rho_\ell^{TE}=C_\ell^{TE}/\sqrt{C_\ell^{TT}C_\ell^{EE}}$, so that the temperature-polarization correlation coefficient is modified by
\beq
\Delta\rho_\ell^{TE} = \int F_X(z) \Delta x_e(z)\,\frac{dz}{1+z}.
\eeq
To reach sampling variance accuracy over some range $\Delta\ell$, each of $\Delta C_\ell^{TT}/C_\ell^{TT}$, $\Delta C_\ell^{EE}/C_\ell^{EE}$, and $\rho_\ell^{TE}$ must be calculated to an accuracy of order $1/\sqrt{\ell\Delta\ell}$, which in the damping tail is of order $3\times 10^{-4}f_{\rm sky}^{-1/2}$.  Since helium recombination lasts for roughly 0.5 $e$-folds, it follows that the accuracy requirement on $x_e(z)$ is roughly $6\times 10^{-4}F^{-1}f_{\rm sky}^{-1/2}$.  In practice, foregrounds (mostly extragalactic) and beam uncertainty will limit the accuracy of small-scale CMB experiments and ``Fisher matrix'' accuracy on the high multipoles may not be reached, nevertheless this does provide a target for the theoretical calculations.

\begin{figure}
\includegraphics[angle=-90,width=3.2in]{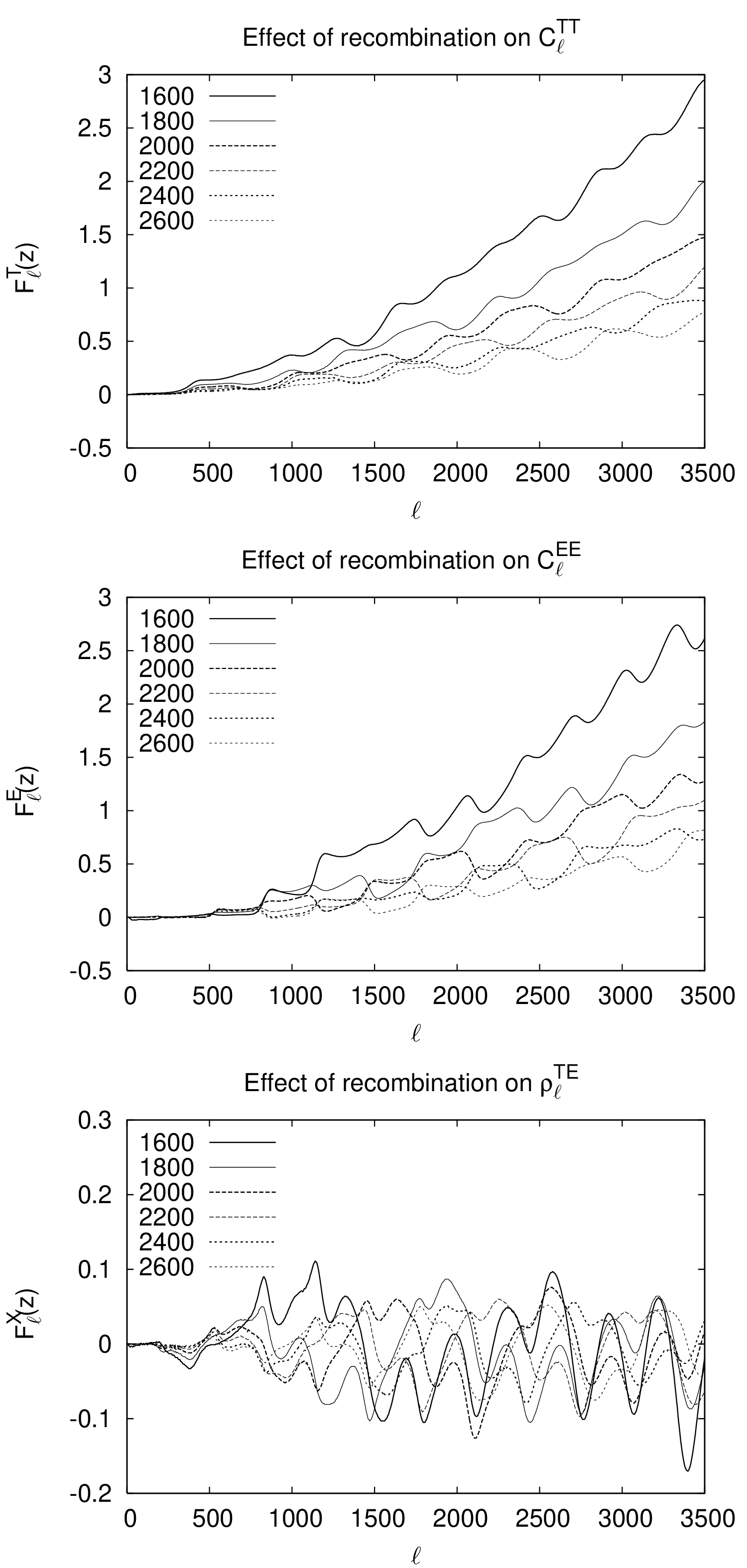}
\caption{\label{fig:ftex}The functions $F_\ell^{T,E,X}(z)$ that describe how recombination influences the CMB power spectrum, where individual contours are for the redshifts specified.}
\end{figure}

The functions $F_\ell^{T,E,X}(z)$ are displayed in Fig.~\ref{fig:ftex}. We have measured these by numerical differentiation of the power spectra 
computed using the {\sc cmbfast} code \cite{1996ApJ...469..437S}.  At each redshift, a Gaussian of width $\sigma_z=100$ in redshift and normalization 
$\int \Delta x_e\,dz=50$ was artificially added to the recombination history, and the change in $C_\ell$ was used to estimate $F_\ell^{T,E,X}(z)$.  
(This is equivalent to taking the functional derivative for sharp test functions and convolving by a Gaussian, and ensures the stability of the CMB 
model calculation.)  Fig.~\ref{fig:ftex} shows several major effects.  One is the trend that $F_\ell^T(z)$ rises from near zero at low multipoles to 
some positive value of order 1--3 for high multipoles, and is most significant for corrections to $x_e(z)$ during later periods of \HeI\ recombination.
This is simply the result of a change in the Silk damping scale: the low multipoles are 
determined by low-$k$ Fourier modes in cosmological perturbation theory, for which the baryon-photon fluid is nearly ideal at these redshifts.  The 
high multipoles are of course Silk-damped, and the sign of $F$ is positive because increasing the electron abundance reduces the photon mean free path 
and hence reduces the diffusion length, thus there is less suppression of power and $C_\ell$ goes up.  The variation in $x_e(z)$ becomes more important 
at late times in \HeI\ recombination because of the increasing overlap with the visibility function.  Indeed, this begins to give an indication of how important a 
solid understanding of \HI\ recombination (which begins $z \sim 1700$) will be.

For polarization, we see that $F_\ell^E(z)\approx F_\ell^T(z)$, because the suppression of acoustic oscillations by Silk damping has roughly the same 
fractional effect on CMB temperature and on polarization.  (While it is true that some polarization is generated during helium recombination due to the 
finite photon mean free path and consequent production of a quadrupole moment, the high Thomson optical depth at that epoch guarantees that this 
polarization is erased before it reaches the observer.)  This is also why the dimensionless correlation coefficient depends very weakly on 
recombination: $|F_\ell^X(z)|\ll|F_\ell^{T,E}(z)|$.

\section{Summary of results for helium, discussion}
\label{sec:summary}

\begin{figure}
\includegraphics[width=3.2in]{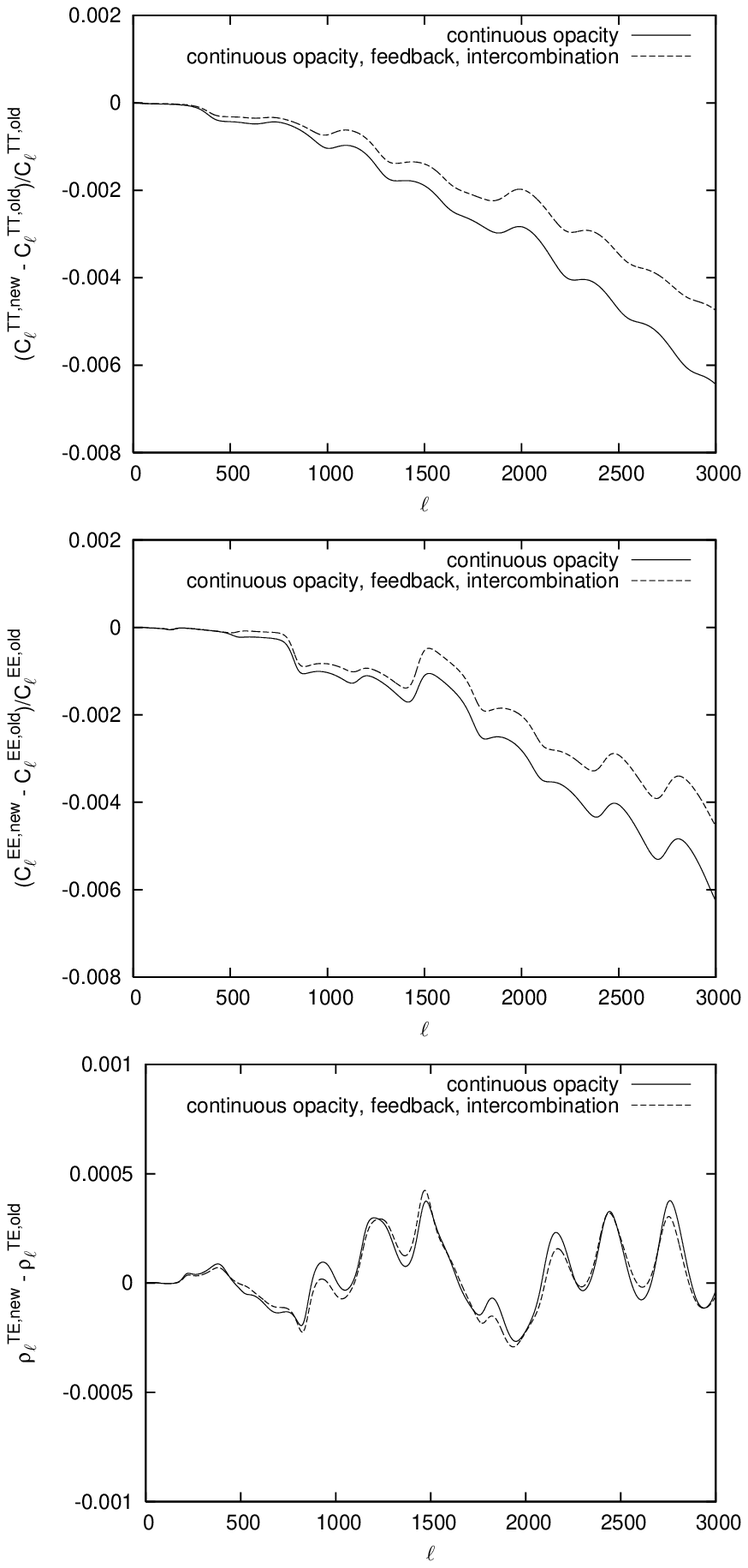}
\caption{\label{figs:cumcll} The cumulative effect on temperature and polarization anisotropies from \HI\ continuous opacity and feedback during \HeI\ recombination, as calculated by {\sc CMBfast}.  We note that here we only consider the difference between the reference helium model with and without these effects.  Comparison of the full \HI\ and \HeI\ history to standard methods such as {\sc Recfast} will be the subject of later work.}
\end{figure}

\begin{table*}
\caption{\label{tab:table1}Summary of the magnitude of effects described in Papers I, II and III.  From the top category to the bottom, we distinguish 
the magnitude of an effect (the systematic error of not including it) from the uncertainty in an effect, and the uncertainty in the implementation.  
Unsigned upper bounds on $| \Delta x_e|$ are indicated by $\pm$.  Note that these are only meant to give an order-of-magnitude bound on the effect, 
where more detail is available in the section cited.  The effect of forbidden processes includes \HI\ opacity in their transport and feedback.  The 
direction shown is either $+$ (increases $x_e$) or $-$ (decreases $x_e$), and a $\sim$ indicates that only an order-of-magnitude was necessary to find 
that the effect was negligible.}
\begin{tabular}{lcccrcccccccccc}
\hline\hline
Effect & & Dir. & & $|\Delta x_e|_{\rm max}\;\;\;$ & & Sec. & & $z$  \\
\hline
Systematic corrections due to effects: &&&&&& \\
\hline
opacity within lines (\LyaHe) & & $-$ & & $2.5 \times 10^{-2}$ & & I, IIIE & & $\sim 1800$ \\
feedback between \LyaHe\ and \InteraHe\ & & $+$ & & $1.5 \times 10^{-2}$ & & I, IIC & & $1800-2600$ \\ 
continuum opacity modification to feedback & & $-$ & & $5 \times 10^{-3}$ & & I, IIIB & & $\sim 1800$  \\
\InteraHe\ inclusion & & $-$ & & $3 \times 10^{-3}$ & & III, \ref{ss:rare} & & $\sim 1900$ \\
\InterHe, $n\ge 3$ inclusion & & $\pm$ & & $ \sim 4 \times 10^{-5}$ & & III, \ref{ss:rare} & & $\sim 2000$ \\
\QuadHe\ inclusion & & $\pm$ & & $\sim 3 \times 10^{-4}$ & & III, \ref{ss:rare} & & $\sim 1900$ \\
opacity in \LyHe\ and \InterHe\ (for $n\ge3$), \QuadHe\ & & $-$ & & $ 5 \times 10^{-4}$ & & I, IIIE & & $\sim 1900$ \\
coherent scattering in \LyaHe\ & & $-$ & & $ 2 \times 10^{-4}$ & & I, IIIE & & $\sim 2000$ \\
distortion, thermal stimulated two-photon effects & & $+$ & & $ 4 \times 10^{-5}$ & & II, II & & $2000-3000$ \\
electron scattering & & $\pm $ & & $3 \times 10^{-4}$ & & III, \ref{sec:thom} & & $1800-2800$ \\
\hline
Uncertainty in the effect's magnitude: &&&&&& \\
\hline
finite linewidth in \HeI\ & & $\pm$ & & $ \sim 4 \times 10^{-4} $ & &  II, VC & & $1800-3000$ \\
non-resonant two-photon effects from $n>2$ & & $\pm$ & & $\sim 5 \times 10^{-4}$ & & II, IV & & $\sim 2000$ \\ 
$\pm 50 \%$ \InteraHe\  spontaneous rate & & $\pm$ & & $\sim 10^{-3}$ & & III, \ref{ss:rare} & & $\sim 1900$ \\
\hline
Uncertainty from the numerical implementation: &&&&&& \\
\hline
modified escape probability grid refinement & & $+$ & & $ \sim 5 \times 10^{-4}$ & & III,\ref{ss:convergence}  & & $\sim 1900$ \\
Monte Carlo resampling & & $\pm$ & & $ \sim 2 \times 10^{-4}$ & & III, \ref{ss:convergence} & & $\sim 1900$ \\
Monte Carlo sample size doubling & & $\pm $ & & $\sim 10^{-4}$ & & III, \ref{ss:convergence} & & $\sim 1900$ \\
level code: half step size & & $\pm$ & & $ \sim 3 \times 10^{-5}$ & & III, \ref{ss:convergence} & & $\sim 1800$ \\
convergence in the modified \LyHe\ series modified $P_{esc}$ & & $-$ & & $\sim 3\times 10^{-4}$ & & I, III E & & $\sim 1900$ \\
$\nmax = 45$ relative to $\nmax = 100$ & & $+$ & & $4\times 10^{-5}$ & & III, IV E & & $\sim 2300$ \\
\hline\hline
\end{tabular}
\end{table*}

\begin{table*}
\caption{\label{tab:summary}Numerical summary of \HeI\ recombination in a model with continuous opacity and feedback between levels.  We provide this for 
comparison with external recombination codes. The cosmological parameters used throughout are, $\Omega_B = 0.04592$, $\Omega_M = 0.27$, 
$\Omega_R=8.23 \times 10^{-5}$, $T_{\rm CMB} = 2.728$, $h=0.71$, $f_{\rm He}=0.079$ and zero curvature.}
\begin{tabular}{cccc|cccc|cccc|cccc|cccc}
\hline\hline
$z$ && $x_e$ && $z$ && $x_e$ && $z$ && $x_e$ && $z$ && $x_e$ && $z$ && $x_e$ \\
\hline
~~~~3087.5 && 1.0790~~~~&& ~~~~2657.3 && 1.0757~~~~&& ~~~~2310.0 && 1.0639~~~~&& ~~~~2008.1 && 1.0413~~~~&& ~~~~1745.6 && 0.9999~~~~\\
~~~~2996.2 && 1.0789~~~~&& ~~~~2604.7 && 1.0743~~~~&& ~~~~2264.2 && 1.0617~~~~&& ~~~~1968.3 && 1.0349~~~~&& ~~~~1711.0 && 0.9994~~~~\\
~~~~2936.9 && 1.0788~~~~&& ~~~~2553.1 && 1.0728~~~~&& ~~~~2219.4 && 1.0594~~~~&& ~~~~1929.3 && 1.0274~~~~&& ~~~~1677.1 && 0.9988~~~~\\
~~~~2878.7 && 1.0787~~~~&& ~~~~2502.5 && 1.0711~~~~&& ~~~~2175.4 && 1.0569~~~~&& ~~~~1891.1 && 1.0196~~~~&& && \\
~~~~2821.7 && 1.0783~~~~&& ~~~~2452.9 && 1.0694~~~~&& ~~~~2132.3 && 1.0539~~~~&& ~~~~1853.6 && 1.0121~~~~&& && \\
~~~~2765.8 && 1.0778~~~~&& ~~~~2404.3 && 1.0677~~~~&& ~~~~2090.1 && 1.0505~~~~&& ~~~~1816.9 && 1.0057~~~~&& && \\
~~~~2711.0 && 1.0769~~~~&& ~~~~2356.7 && 1.0658~~~~&& ~~~~2048.7 && 1.0464~~~~&& ~~~~1780.9 && 1.0015~~~~&& && \\
\hline \hline
\end{tabular}
\end{table*}

\begin{figure}
\includegraphics[width=2.3in, angle=-90]{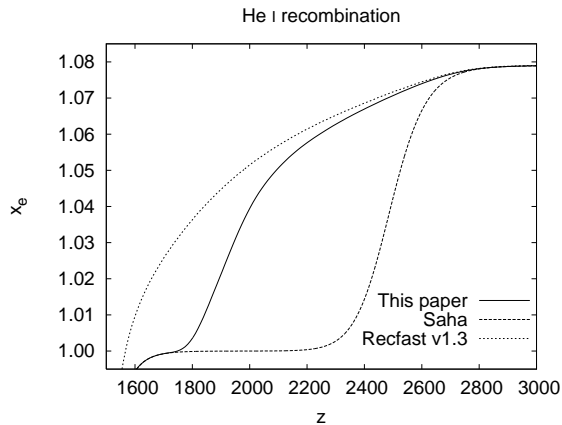}
\caption{\label{figs:heirecomb} The \HeI\ recombination history from our multi-level atom code (solid line), compared to the Saha equation (long-dashed) 
and the commonly used three-level code {\sc recfast} by Seager et~al. \cite{1999ApJ...523L...1S} (short-dashed).  Both our analysis and that of Seager et~al. find 
that \HeI\ recombination is delayed due to the $n=2$ bottleneck.  However we find a slightly faster recombination than Seager et~al. due primarily to our inclusion 
of the intercombination line \HeI] $2^3P^o-1^1S$ and the accelerating effect of \HI\ opacity.  The latter effect causes our \HeI\ recombination to finish at $z\approx 
1700$, whereas in {\sc recfast} one-third of the helium is still ionized at that time.}
\end{figure}

\begin{figure}
\includegraphics[width=2.3in, angle=-90]{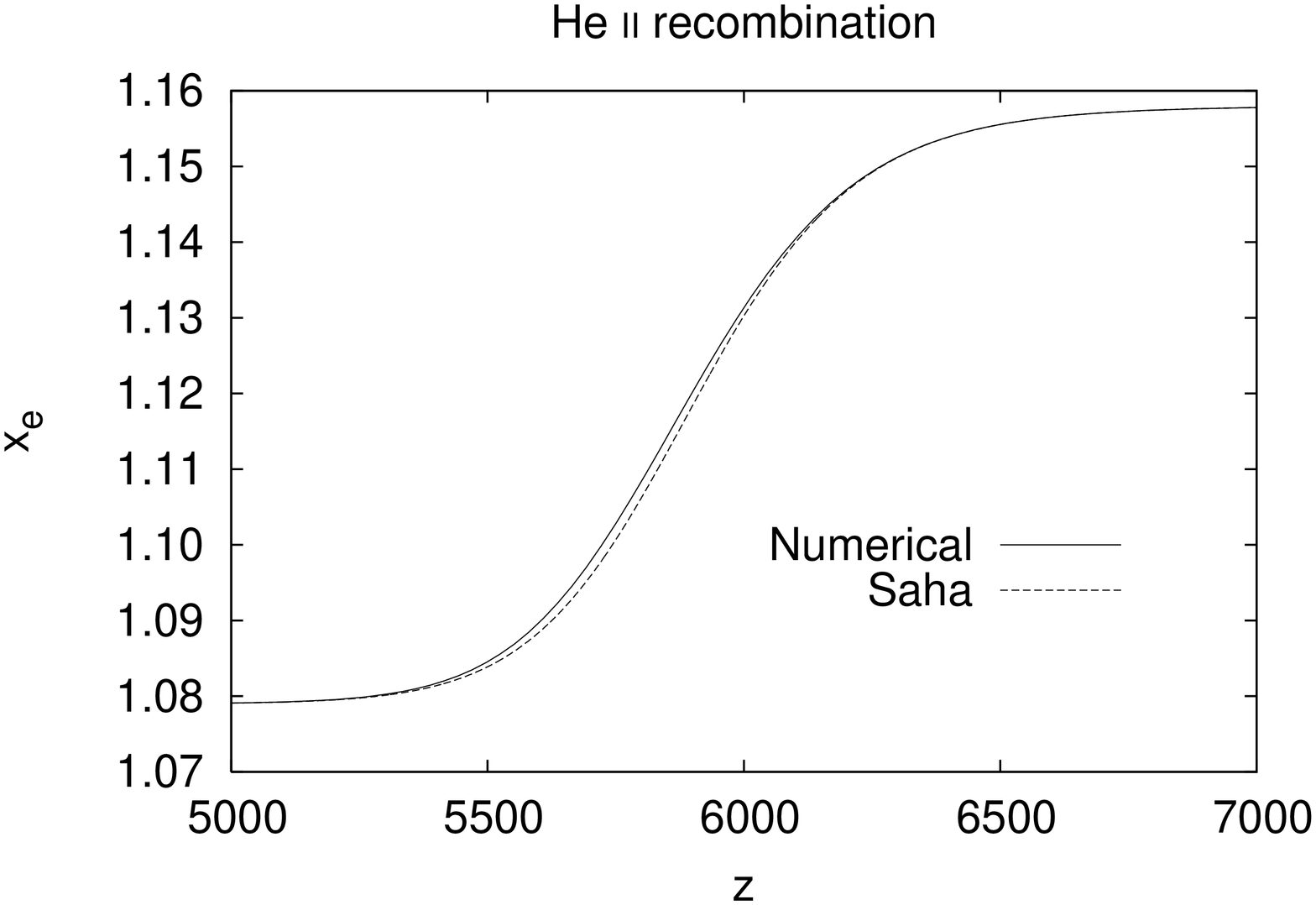}
\caption{\label{figs:heiirecomb} The \HeII\ recombination history from the level code developed here.  \HeII\ recombination is essentially irrelevant for CMB physics, 
and because of its rates it varies from Saha evolution at the level of $< 0.2\%$.  (for example, at $\ell < 3000$ the absolute difference in $C_\ell^{\rm TT}$ for the full 
model relative to Saha is $< 3 \times 10^{-5}$.)}
\end{figure}

In the series of Papers I-III, we have addressed a multitude of effects in \HeI\ recombination, some new, and some that have appeared recently in 
literature \cite{2006A&A...446...39C, 2005AstL...31..359D, 2004MNRAS.349..632L, 2006AstL...32..795K}.
Here, we described the effect of Thomson and \Hethree\ scattering, 
peculiar velocities in the recombination plasma, and collisional and rare processes.  Table~\ref{tab:table1} summarizes the processes described in 
Papers I-III and the magnitudes of their effects, and Fig.~\ref{figs:cumcll} shows the cumulative difference in the CMB temperature and polarization 
anisotropies due to the largest corrections described.  We want to emphasize only the modification to the \HeI\ recombination history from the effects 
described in Papers I-III, so we only compare with a reference model where those effects are absent.  A summary of the final \HeI\ recombination history
developed here can be found in Fig.~\ref{figs:heirecomb}.  We have not described \HeII\ recombination in detail because it 
is almost a pure Saha recombinaton, as shown in Fig.~\ref{figs:heiirecomb} and is insignificant for the CMB anisotropy.  

Thomson and \Hethree\ scattering contribute to $x_e$ only at the level of a few $\times 10^{-4}$ and can be neglected (for comparison, the uncertainty 
associated with the interpolation of the $P_{\rm esc}$ derived in Monte Carlo used here is only of order $10^{-4}$).  The largest (known) systematic 
error presented here is from plausible uncertainties \cite{1977PhRvA..15..154L, 1969ApJ...157..459D, 1978JPhB...11L.391L} in the \InteraHe\ rate and is 
of order $|\Delta x_e| \sim 10^{-3}$.  Making no assumptions about the correlations between systematic errors, a conservative estimate of the overall systematic in 
$x_e(z)$ from known effects is $3 \times 10^{-3}$.

The remaining uncertainty in the $C_\ell$s due to helium recombination can be estimated by multiplying the maximum error in the electron abundance, $|\Delta x_e|_{\rm max}=0.003$,
by $\int |F^T_\ell(z)|\,d\ln(1+z)$ (see Eq.~\ref{eq:ftz}).  Over the redshift range of helium recombination, $1700<z<2800$, we find $\int |F^T_\ell(z)|\,d\ln(1+z)=0.6$ at $\ell=3500$, implying that $C_\ell^{TT}$
has a fractional uncertainty of 0.002 (this number is lower for smaller $\ell$).  The same calculation for $C_\ell^{EE}$ also gives 0.002, and for $\rho_\ell^{TE}$ we get $<3\times 10^{-4}$ at $\ell<3500$.
The latter number is roughly equal to the cosmic variance limit $\sim \ell_{\rm max}^{-1}$.  Therefore we believe that our present calculation of helium recombination is sufficient to predict, at $\ell<3500$, the
temperature and $E$-mode polarization power spectra to 0.2\% accuracy and the correlation coefficient $\rho_\ell^{TE}$ to cosmic variance accuracy.  Note that this uncertainty does {\em not} include the
effects of hydrogen recombination.  Also, while we have made every effort to include all effects, we acknowledge that there could be additional processes that escaped our imagination.  We encourage further work to ensure the 
completeness of the understanding of \HeI\ recombination.

We conclude that a modification yielding more accurate \HeI\ recombination histories will give a negligible change in the WMAP results.  However, the 
considerations presented here are significant for the next generation of small-scale anisotropy experiments, which seek to measure the large multipoles 
at the percent level.  The changes shown in Fig.~\ref{figs:cumcll} would affect a cosmic variance limited experiment at the $\sim 1\sigma$ level at 
$\ell_{\rm max}=1500$, and $\sim 8\sigma$ at $\ell_{\rm max}=3000$.  This means that the effect is in principle at the 
$\sim 1\sigma$ level for the upcoming Planck satellite \cite{2006MNRAS.373..561L} or (nearly so) for a high-resolution CMB experiment mapping 1--2\% of the sky to $\ell_{\rm 
max}=3000$, although it is possible that systematic uncertainties such as beam modeling and point source removal may prevent one from reaching ``Fisher 
matrix'' accuracy.  Of the corrections considered, we propose that feedback of non-thermal distortions between the allowed lines, continuum opacity 
from \HI\ photoionization, and the inclusion of the \InteraHe\ \cite{2005AstL...31..359D} rate become part of the standard recombination model.  We also 
emphasize that one of the largest uncertainties in the \HeI\ recombination treatment presented here is the \InteraHe\ spontaneous rate.
In Paper II we show that the corrections due to two-photon \cite{2005AstL...31..359D} and finite linewidth effects are almost entirely negligible.  A fast, modified 
recombination code based on these corrections will be the subject of later work.

While \HeI\ recombination is important cosmologically, it should only be considered as a first step toward a much broader treatment of cosmological recombination as a whole.  
This includes \HI\ recombination, which is especially important for the CMB since it affects the free electron abundance near the peak of the CMB visibility function.  (It will also 
affect the temperature evolution, 21 cm absorption, and chemistry during the ``Dark Ages'' between recombination and reionization.)  The calculation of \HI\ recombination and 
its effects on the $C_\ell$s to sub-percent accuracy will be far more difficult than the calculations for \HeI\ presented in this series of papers.  One reason is that there are 12 H 
atoms for every He atom, so to achieve the same absolute accuracy in $x_e$ requires a much better accuracy in the ionization state of H than of He.  Another reason is that the 
optical depth in the \HI\ Ly$\alpha$ line can be of order $10^9$ (instead of $10^7$ for the analogous 21.2 eV line of \HeI), which means that slow processes can play a larger 
role, and that the finite linewidth effects could potentially be much more significant.  (The high optical depth also makes the photon Monte Carlo developed here computationally 
demanding for hydrogen.)  Finally, the very high-$n$ states of \HI\ become important at lower temperatures, and both $l$-dependent occupations and collisional processes in 
these states will have to be carefully taken into account \cite{2007MNRAS.374.1310C}.  It is because of these complexities that we have not solved \HI\ recombination here; rather 
we plan to follow up this series of papers with a similar series on hydrogen in the future.

\begin{acknowledgments}

E.S. acknowledges the support from grants NASA LTSAA03-000-0090 and NSF PHY-0355328.  C.H. is a John Bahcall Fellow in Astrophysics at the Institute for 
Advanced Study.  We acknowledge useful conversations with Jens Chluba, Bruce Draine, Jim Peebles, Doug Scott, Uro\v s Seljak, and Rashid Sunyaev.  
We also thank Joanna Dunkley for critical readings and comments prior to publication.

\end{acknowledgments}
\appendix

\section{The electron scattering kernel characteristic for dipole angular distributions}
\label{sec:appcharele}

In this section, we solve explicitly for the characteristic function of the electron scattering distribution over frequency ($P(\Delta \nu)$ taken to be in the Thomson limit),
\beq \varpi(k) = \langle e^{ik(\Delta \nu)} \rangle_{\Delta \nu}. \eeq
This is used in the Fourier domain solution to the transport equations for complete redistribution with electron scattering and continuum opacity which is negligible over
the width of the line, described in Sec.~\ref{ss:qintegral} and is applicable for the intercombination and quadrupole lines for $z > 2000$.  Substituting the variance 
$\zeta=2 \sigma_D^2 [1 - \cos(\chi)]$ and inserting the dipole angular redistribution function, this becomes
\beqa \varpi(k) &=& \frac{3}{16 \sigma_D^2} \int_0^{4 \sigma_D^2} \left [ \frac{\zeta^2}{4 \sigma_D^4} - \frac{\zeta}{\sigma_D^2} + 2 \right ]
e^{-k^2\zeta/2} d\zeta \nonumber \\ 
&=&  \frac{3}{16 \sigma_D^2} \left [ \frac{R_1(k)}{4 \sigma_D^4} - \frac{R_2(k)}{\sigma_D^2} + 2 R_3(k) \right ], \eeqa
where we have absorbed the Gaussian integrals $R_1$, $R_2$, and $R_3$ which are explicitly,
\beqa
R_1(k) &=& \frac{16}{k^6} \left \{ 1 - \left [ 2 \sigma_D^4 k^4 + 2 \sigma_D^2 k^2 + 1 \right ] e^{-2 k^2 \sigma_D^2} \right \} \nonumber \\
R_2(k) &=& \frac{4}{k^4} \left [ 1 - \left ( 2 \sigma_D^2 k^2 +1 \right ) e^{-2 k^2 \sigma_D^2} \right ]\nonumber \\
R_3(k) &=& \frac{2}{k^2}\left ( 1 - e^{-2 k^2 \sigma_D^2} \right ).
\eeqa
Note that for an isotropic distribution, the result can be expressed very concisely as $\varpi(k) = (4 \sigma_D^2)^{-1} R_3(k)$.
Note that these require caution numerically for small $k$ because of the near-cancellation of several terms.  This can be remedied 
by using a high-order expansion in small $k$ (the first order cancels and leaves divergent terms in $k \rightarrow 0$).  We also note that while the relativistic electron scattering 
kernel is in general asymmetric about $\Delta \nu=0$, in the Thomson scattering approximation considered here the kernel is symmetric, leading to all-real $\varpi(k)$.

\section{Fluid description of the baryons}
\label{app:cpt}

In this appendix, we investigate whether the treatment of the baryons as a single fluid is adequate for the investigation of the peculiar velocity
field.  In particular, we would like to understand whether the single fluid treatment accurately describes Silk damping.  To do this, we have to
determine whether the collisional relaxation rates are fast compared to the oscillation timescale at the Silk damping length, $k_D^{-1}$.  We will show in this
appendix that the collisional relaxation rates are indeed fast, so it should be accurate to describe the baryons as a fluid.  Silk damping then proceeds
as in the usual picture at all scales $(k_D \sim 0.37~{\rm Mpc}^{-1}) <k < (k_{\rm fs} \sim 2~{\rm Mpc}^{-1})$ at $z=2000$ (with smaller scales 
evolving in this range at earlier times \cite{2005MNRAS.363..521G}).  Below the photon free-streaming 
scale (the mean free path of photons), only electromagnetic interactions can influence the baryon velocity moment.  
We will consider the possibility of charge separation on these scales and show that it is a negligible contribution to the velocity structure. 

There are five constituents of the baryonic plasma during the recombination epoch: the electrons, ions (H$^+$, He$^+$, He$^{2+}$), and neutral
species (He).  (During helium recombination there is very little neutral H, and its velocity structure is not relevant because its only significant
role is to provide continuum opacity.)
These species acquire and exchange momentum through several process \cite{2001NewA....6...17H}:
(1) radiation pressure, which is only significant for the electrons, by mass,
(2) electric fields due to charge separation,
(3) collisions, and
(4) photoionization/recombination, which can switch particles between the He and He$^+$ constituents.
In principle one should check energy exchange rates (e.g. whether the electrons and ions thermalize to a common temperature), however since the baryon
thermal pressure plays no role in Silk damping these are not relevant here.

Siegel \& Fry \cite{2006ApJ...651..627S} have considered electric fields due to charge separation and find that they efficiently prevent the electron
and ion densities from departing significantly from each other.  In particular, during recombination ($z\sim $few$\times 10^3$) they find
[See \cite{2006ApJ...651..627S}, Eq.~(24)]
\beq
\label{eqn:Siegel}
\theta_q\equiv \theta_i-\theta_e = -\frac{\sigma_{\rm T}m_p c}{3\pi e^2}\frac{\rho_\gamma}{\rho_b}\frac{d}{dt}(\theta_\gamma-\theta_b);
\eeq
in their analysis the ions ($i$) consisted entirely of protons but the addition of some \HeII\ or \HeIII\ should cause no qualitative changes.
(It should result in the ion expansion $\theta_i$ being replaced by the charge-weighted expansion of all the positive ions, $\sum_j q_j\theta_j 
n_j/n_e$, where $q_j$ is ion charge and $n_j/n_e$ is a number density ratio.)
This equation results from a balance of the radiation pressure on the electrons, which is the driving term in their separation from the ions, with the
electrostatic force that seeks to eliminate bulk charges.  Here
$\theta$ is the peculiar expansion, $\sigma_{\rm T}$ is the Thomson cross section, and $\rho_\gamma/\rho_b$ is the ratio of photon to baryon densities.
Plugging in the cosmological parameters gives
\beq
\theta_q = - \Delta t_q\;\frac{d}{dt}(\theta_\gamma-\theta_b).
\eeq
where
\beq
\Delta t_q = 6\times 10^{-20}\,{\rm s}\,\left(\frac{1+z}{2000}\right).
\eeq
During an acoustic oscillation $\theta_\gamma$ and $\theta_b$ are similar, but both on the order of $10\,$km$\,$s$^{-1}$ or less.  Thus so long as
$\Delta t_q$ is much shorter than the oscillation period, $\Delta t_q\omega\ll 1$, the electron and ion velocities will be similar.

\begin{figure}
\includegraphics[width=3.2in]{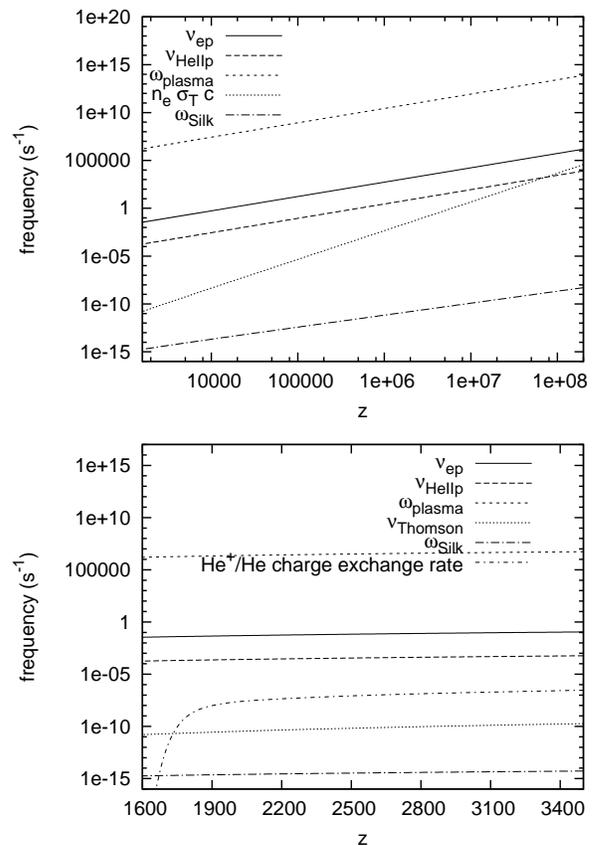}
\caption{\label{fig:scales}  Several scales in the recombination plasma relevant for helium peculiar velocities.  The fastest
here is the plasma frequency, followed by the electron-proton (and then \HeII-p) momentum transfer rate.  Note that the plasma frequency is
significantly faster than both the Thomson rate and the frequency of an acoustic oscillation at the Silk scale.  This
greatly suppresses the magnitude of the charge separation, see Eq.~\ref{eqn:Siegel}.  In the lower plot, we focus on the 
region neutral helium evolution ($z<3500$).  The charge exchange momentum transfer rates between ${\rm He}$ and ${\rm He}^+$ are 
much larger than the frequency of baryon acoustic oscillations at the Silk scale during the period of neutral helium recombination.  
This brings neutral helium into a common fluid with the other charged baryons.}
\end{figure}

Next we come to the collisional momentum exchange times.  Of interest are the electron-ion collision rate, the collision rates between ions of various
species, and the neutral-ion collision rate.
The collisional momentum relaxation rate for particle type $1$ against particle type $2$ (where we will take $1$, and $2$ to be electron-proton,
and \HeII-proton) for a thermal distribution and small differential drift velocity is given by
\beq
\nu_{12} = \frac{2}{3 \sqrt{ 2 \pi}} n_2 \left ( e^2 Z_1 Z_2 \right )^2 \frac{4 \pi}{\mu m_1 v_\mu^3} \ln (\Lambda)
\eeq
where $\Lambda$ is the Coulomb logarithm, $\mu = m_1 m_2 / (m_1+m_2)$ is the reduced mass, and $v_\mu = \sqrt{k_B T/\mu}$ \cite{1952ApJ...116..299S}.
For roughly $z>6000$, the doubly-ionized helium population exceeds the singly-ionized population.  The momentum relaxation rate for
doubly-ionized helium against protons is four times larger than the \HeII-proton rate.  
For simplicity, we will only consider the \HeII-proton rate in Fig.~\ref{fig:scales},
for, if it is sufficient to relax \HeII\ to the velocity structure of the other baryons, the \HeIII-proton rate must also be sufficient.  Note also
that the relaxation rate for \HeII\ on protons has no dependence on the number density of \HeII, while the relxation rate for protons on \HeII\
does.

We will consider only proton collisions with singly-ionized helium, even though at for $z>6000$, the  
The rates involving neutral He must be considered at $z<3500$ when \HeI\ is present; they are usually slower because they
lack the long-range nature of the Coulomb force.  The dominant rate is that of resonant charge exchange with He$^+$:
\begin{equation}
{\rm He} + {\rm He}^+ \rightarrow {\rm He}^+ + {\rm He}.
\end{equation}
The charge exchange momentum transfer rate is well-approximated by \cite{1966PSS...14.1105B}
\begin{eqnarray}
\nu_{{\rm He}^+,{\rm He}} &=& 4.4 \times 10^{-13} n_{{\rm He}^+} (2T_{\rm m})^{1/2} \nonumber \\
&\times& (11.6 - 1.04 \log_{10}(2T_{\rm m})) {\rm s}^{-1}, 
\end{eqnarray}
where $T_{\rm m}$ is in Kelvins and $n_{{\rm He}^+}$ is in cm$^{-3}$.
We have not considered momentum exchange between He and He$^+$ due to photoionization/recombination, or He-proton scattering;
if we did then this would only strengthen the conclusion that the momentum exchange rate to the charged fluid components is fast.

The acoustic oscillation frequency at the Silk damping length is given by
\begin{equation}
\omega_{\rm D} = \frac{ck_{\rm D}}{\sqrt{3(1+R)}},
\end{equation}
where the damping wavenumber $k_{\rm D}$ is obtained as in Ref.~\cite{1995PhRvD..52.3276Z}.  The
damping length $k_{\rm D}^{-1}$ runs from $10^{-7}\,$Mpc  
comoving at $z=2\times 10^8$ to $3\,$Mpc at $z=1600$.  (Before $z=2\times 10^7$
the usual computation is not valid because electron-positron pairs increase the opacity.)

We have plotted the momentum exchange rates and compared them to the plasma frequency $\omega_p$ and the acoustic oscillation frequency at the Silk
damping length $\omega_{\rm D}$ in Fig.~\ref{fig:scales}.  The plot runs from $z=2\times 10^8$ until $z=1600$ when helium recombination has completed, for
practical purposes.  In all cases the relevant collisional rates are many orders of
magnitude faster than the acoustic oscillations.  This means that for wavenumbers $k^{-1}>10^{-7}\,$Mpc we expect that hydrodynamics is valid, Silk
damping will occur, and the exponential suppression of velocity perturbations predicted by the usual treatment is correct.

A more detailed treatment is required in order to understand what happens at scales below $10^{-7}\,$Mpc comoving.  Our physical expectation is that
acoustic oscillations at such scales would also be damped by photon diffusion, with the Silk damping slightly modified by inclusion of positrons, and
by the analogous process of neutrino diffusion (since these scales are well within the horizon when neutrinos decouple).  Even if this does not happen,
the photon mean free path at $z\sim 2\times 10^8$ is $10^{-10}\,$Mpc comoving, so for $k\le 10^{10}\,$Mpc$^{-1}$ the usual Silk damping calculation
applies; since these scales are smaller than the Silk damping length they will be exponentially damped.  For higher $k$ it is less obvious what
happens, but since the neutral He mean free path is much larger than $10^{-10}\,$Mpc comoving during He recombination there can be no structure in the
neutral He peculiar velocity field at smaller scales.

\bibliography{recombination3}
\end{document}